\documentstyle[12pt,kay_bib,a4,epsf,amssymb]{article}
\renewcommand{\theequation}{\arabic{section}.\arabic{equation}}

   \newcommand{\e}{\mbox{\large e}}

   \newcommand{\ggl}{\raisebox{-0.5ex}[0em][-0.5em]{$\stackrel{\Sc 
                      >}{\Sc \sim}$}}
   \newcommand{\kgl}{\raisebox{-0.5ex}[0em][-0.5em]{$\stackrel{\Sc 
                      <}{\Sc \sim}$}}

   \newcommand{\Sc}{\scriptstyle}
   \newcommand{\dis}{\displaystyle}
   \newcommand{\tex}{\textstyle}
   \newcommand{\be}{\begin{equation}}
   \newcommand{\ee}{\end{equation}}
   \newcommand{\ba}{\begin{eqnarray}}
   \newcommand{\ea}{\end{eqnarray}}

\begin{document}
\title{\vspace*{-3cm}
\Large\bf
Diffusion of a Test Chain in a Quenched Background of Semidilute Polymers}
\author{
   {\sc Stefan M\"uller}
  \\[1.5ex]
  \it Fachbereich Physik der Universit\"{a}t Essen
  \\
  \it D-45117 Essen, Germany
  \\
  \it E-Mail: smull@next17.theo-phys.uni-essen.de}

\maketitle
\thispagestyle{empty}
\thispagestyle{empty}
\begin{abstract}
{\small 
Based on a recently established formalism (U. Ebert, {\em J.\ Stat.\ Phys.} {\bf 82} (1996) 183: \cite{utestat}) we analyze the diffusive motion of a long polymer in a quenched random medium. The medium is modeled by a frozen semidilute polymer system. In the framework of standard renormalization group (RG) theory we present a systematic perturbative approach to handle such a many chain system. In contrast to \cite{utestat} we here deal with long range correlated disorder and find an attractive RG fixed point. Unlike in polymer statics the semidilute limit here yields new nontrivial power laws for dynamic quantities. The exponents are intermediate between the Rouse and reptation results. An explicit one loop calculation for the center of mass motion is given.\vspace*{1cm}
 }  
\end{abstract}

\pagenumbering{arabic} 
\setcounter{section}{0}
\setcounter{page}{1}  

\section{Introduction}

The diffusion of a polymer chain in a quenched environment of obstacles still poses many open questions. It is of direct relevance for the motion of a chain through a gel. Theoretically two opposit approaches have been pursued: The reptation picture in its original form treats the environment as an ordered lattice of impenetrable obstacles \cite{DeGerept,DoiEdw}. It neglects all interaction effects except for the impenetrability and it ignores the disordered character of the environment, which may lead to entropic trapping \cite{MuBa89}. In the asymptotic limit of long chains it predicts simple power laws. The opposite approach models the environment as a Gaussian distributed random potential with correlations of microscopic range \cite{utestat,Mamemo81,Mach89,EbBaSch}. It thus concentrates on energetic effects, and, in a coarse grained sense, also on entropic traps. It however completely neglects the impenetrability of the obstacles and it ignores all correlations in the quenched environment. This approach results in slowing down of the chain motion which is much stronger than predicted by reptation theory. In particular it has been argued that the diffusion coefficient of the chain decreases exponentially with increasing chain length \cite{utestat,Mach89}.

The present work builds upon the random potential model, but includes realistic correlations of the obstacles. The environment is modeled as a quenched equilibrium configuration of a polymer solution, where the chains forming the environment may chemically differ from the mobile chain. In the semidilute limit of strongly overlapping background chains this can be considered as a good model for a gel, except that we ignore the dynamics of short strands of the gel. More serious is the fact that we treat all chains as phantom chains. They interact by excluded volume interaction, but of course the mobile chain can penetrate the strands of the gel.

Our model introduces the macroscopic correlation length $\xi^{(B)}$ of the density correlations of the environment, which can be identified with the blob size $\xi_c$ (concentration blobs) of the quenched semidilute solution of the background chains \cite{DeGe}. Two other important length scales are the radius of gyration $R_g$ of the mobile chain and the average radius of gyration $R_g^{(B)}$ of the background chains. This is in contrast to Gaussian disorder models, where the correlation length of the disorder is of the order of the segment size of the mobile chain, and where no scale corresponding to $R_g^{(B)}$ shows up. 

Our model turns out to be renormalizable, most similar to standard ternary polymer solutions. Renormalization drives the system into fixed points, and in limiting situations power law scaling results. The exponents are intermediate between Rouse type behaviour and reptation results.

We performed an explicit one loop calculation of the center of mass motion,
\be
R^2(t):=\overline{\left\langle{\left({\bf R}_{cm}(t)-{\bf R}_{cm}(0)\right)}^2\right\rangle}\:,
\label{R2first}
\ee
where the angular brackets denote the average over the stochastic process of the test chain motion and the bar stands for averaging over the quenched semidilute background. Switching off the interaction with the random background we simply deal with the free draining limit, which for the center of mass motion yields a diffusive behaviour identical to that of a pure Rouse chain: $R^2(t)\sim t/n$ for all times. Here $n$ denotes the segment number of the moving chain. With excluded volume interaction between the moving chain and the background polymers we find three time regimes:\\
(i) As long as the polymer has moved much less than the blob size of the background, it does not notice the hindering due to the frozen chains. We obtain approximately free diffusion: $R^2(t)=6D_0t$, where $D_0\sim 1/n$, and call this regime the ultra-short time regime.\\ 
(ii) What follows is a short time regime, where the moving polymer interacts with the background chains. This leads to a slowing down. We find that in the case of strongly overlapping chains there is a universal time regime where pure subdiffusive power law behaviour emerges: $R^2(t)\sim t^{.78}$.\\
(iii) For very long times one always finds diffusive behaviour, $R^2(t)=6Dt$, but the diffusion coefficient is reduced as compared to Rouse behaviour. In the semidilute limit the diffusion constant decreases with chain length resp.\ monomer concentration as $D\sim c^{-.45}n^{-1.45}$. 
 
Since these exponents are obtained from a first order $\epsilon$-expansion setting $\epsilon=4-d=1$ in three dimensions, their quantitative precision is limited. Due to the technical complexity of the calculations a higher loop evaluation however seems out of reach.

This paper is organized as follows: In sect.\ 2 we formulate the model and set up the dynamic generating functional which in principle allows for a perturbative calculation of any correlation function concerning the mobile chain. In sect.\ 3 we perform a bare one loop calculation of the center of mass motion. Sect.\ 4 is devoted to the renormalization of the model. To renormalize the static parameters of the model we establish a connection to ternary polymer solutions. The dynamic renormalization is examined by means of the one loop result for the center of mass motion. The final scaling results for the center of mass motion are presented in sect.\ 5, followed by a concluding discussion in sect.\ 6. Two appendices are devoted to the relation between quenched and annealed averages in polymer statics and discuss the case of Gaussian polymers, respectively.

\setcounter{equation}{0}
\section{Formalism}
\subsection{The model}
We employ the standard Langevin-dynamics of a discrete chain consisting of $n$ segments:
\be\frac{\partial}{\partial t}{\bf r}_i(t)
 = \gamma \Big(-\frac{\partial}{\partial {\bf r}_i(t)} {{\cal H}_M} + {\bf f}_i(t)\Big),
\label{langev}
\ee
where ${\bf r}_i(t)\in{{\mbox I}\!{\mbox R}}^d$ $(i=1,\ldots,n)$ fixes the position of the $i$th segment of the moving chain in a $d$-dimensional space and ${\partial}/{\partial {\bf r}}$ stands for the gradient. The thermal random forces ${f}_{i\mu}(t)$ $(\mu=1,\ldots,d)$ are modeled by Gaussian white noise. Their second moment is connected to the microscopic mobility $\gamma$ of a segment:
\be
\left\langle f_{i\mu}(t)f_{j\nu}(t')\right\rangle= \frac{2}{\gamma}\delta_{ij}\delta_{\mu\nu}\delta(t-t')\:.
\label{whinoi}
\ee
This guarantees the correct equilibrium distribution ${\cal P}_{eq.}\sim e^{-{\cal H}_M}$ for the chain (we set $k_BT=1$). The Hamiltonian ${\cal H}_M$ consists of three parts:
\ba
\mbox{\hspace*{-9.5 cm}}\lefteqn{{\cal H}_M[{\bf r}_1(t),\ldots,{{\bf r}_n}(t);\{{\bf r}^{(B)}\}] =}
\nonumber
\ea
\be
\frac{1}{4l^2}\sum_{i=2}^n({\bf r}_i(t) -
{\bf r}_{i-1}(t))^2 + u_0 l^d \sum_{1 \leq i < j \leq n}
\delta^{d}({\bf r}_{i}(t) - {\bf r}_j(t)) + v_0l^d\sum_{i=1}^n \rho^{(B)} ({\bf r}_i(t))\:.\nonumber\\
\label{ham}
\ee
The first two parts describe chain connectivity resp.\ excluded volume interaction in a discrete spring and bead model, where the effective segment size is determined by the microscopic length $l$. The last term describes the interaction with the background.

The quenched random background consists of $M$ spring and bead chains with lengths $n_1,\ldots,n_M$. A configuration of the background is given by all segment positions of all chains:
\be
\{{\bf r}^{(B)}\}=\underbrace{{\bf r}^{(B)}_{1,1},\ldots,{\bf r}^{(B)}_{1,n_1}}_{\mbox{\small chain 1}};\underbrace{{\bf r}^{(B)}_{2,1},\ldots,{\bf r}^{(B)}_{2,n_2}}_{\mbox{\small chain 2}};\ldots;\underbrace{{\bf r}^{(B)}_{M,1},\ldots,{\bf r}^{(B)}_{M,n_M}}_{\mbox{\small chain M}}\:.         
\label{segvec}
\ee
The segment density of the background at point $\bf r$ then reads 
\be
\rho^{(B)}({\bf r})= \sum_{m=1}^M\sum_{j=1}^{n_m}\delta^d({\bf r}-{\bf r}_{m,j}^{(B)})\:,
\label{densi}
\ee
i.e.\ the interaction between the moving chain and the background chains in eq.~(\ref{ham}) is nothing else than the excluded volume interaction among segments of the moving chain and segments of the background chains. The corresponding interaction constant is $v_0$, while $u_0$ in eq.~(\ref{ham}) governs the intrachain-interaction among segments of the moving chain. Note that the ${\bf r}_{m,j}^{(B)}$ are independent of $t$.

Finally we have to specify the distribution function ${\cal P}_M^{(B)}$ of the background polymers. We here use a canonical ensemble of chains, just as if the background would be a polymer solution in equilibrium:
\be
{\cal P}_M^{(B)}[\{{\bf r}^{(B)}\}]=\frac{1}{{\cal Z}_M^{(B)}}\: \e^{\tex -{\cal H}_M^{(B)}}\:, 
\label{distrib}
\ee
where
\ba
\lefteqn{{\cal H}_M^{(B)}[\{{\bf r}^{(B)}\}] =} 
\nonumber\\& &
\frac{1}{4l^2}\sum_{m=1}^M\sum_{j=2}^{n_m}({\bf r}_{m,j}^{(B)} -
{\bf r}_{m,j-1}^{(B)})^2 + u_0^{(B)} l^d \sum_{{\mbox{\scriptsize all pairs}}\atop (m_1, j_1);(m_2, j_2)}\delta^d({\bf r}_{m_1,j_1}^{(B)}-{\bf r}_{m_2,j_2}^{(B)})\nonumber\\
\label{backh}
\ea
is the spring and bead Hamiltonian of $M$ chains with an excluded volume interaction constant $u_0^{(B)}$ among all segments of the background and 
\be
{\cal Z}_M^{(B)}=\int{\cal D}\{{\bf r}^{(B)}\}\, \e^{\tex -{\cal H}_M^{(B)}} 
\label{backz}
\ee
denotes the partition function. The integration $\int{\cal D}\{{\bf r}^{(B)}\}$ means summation over all configurations of the background polymers in a $d$-dimensional volume $\Omega\to\infty$, i.e.
\be
{\cal D}\{{\bf r}^{(B)}\}= \prod_{m=1}^M\prod_{j=1}^{n_m}\frac{d^dr_{m,j}^{(B)}}{(4 \pi l^2)^{d/2}}\:. 
\label{meas}
\ee
The factor $(4 \pi l^2)^{d/2}$ is introduced for convenience. While angular brackets denote the average over thermal forces, a bar stands in the following for averaging any observable ${\cal O}$ over the frozen background distribution:
\be
\overline{\cal O}=\int{\cal D}\{{\bf r}^{(B)}\} \,{\cal P}_M^{(B)}[\{{\bf r}^{(B)}\}]\:{\cal O}[\{{\bf r}^{(B)}\}]\:. 
\label{cano}
\ee 
For explicit calculations it is more convenient to switch to the equivalent grand canonical ensemble. Thus all final results are understood to hold in the thermodynamic limit of infinitely many background chains in an infinite volume $\Omega\to\infty$, where the segment concentration
\be
c=\frac{\sum_{m=1}^M n_m}{\Omega} 
\label{segconc}
\ee
is kept finite. A general polydisperse grand canonical polymer ensemble is characterized by the function $c_p(n^{(B)})$, giving the concentration of chains of length $n^{(B)}\in\{1,2,\ldots,\infty\}$. The total chain concentration
\be
c_p=\sum_{n^{(B)}=1}^{\infty} c_p(n^{(B)})
\label{cppol}
\ee
is connected to the segment concentration $c$ via
\be 
c=N c_p\:,
\label{c}
\ee
where $N$ denotes the average chain length of the background chains
\be 
N=\sum_{n^{(B)}=1}^{\infty} n^{(B)}\frac{c_p(n^{(B)})}{c_p}\:.
\label{N}
\ee
In conclusion we should point out that the model presented here is known to include all relevant terms in the RG sense, i.e.\ contributions from any short-ranged modification of the model like three-body interactions vanish in the limit of long chains. As stressed before our model however does not fall into the universality class of short range Gaussian correlated disorder since the quenched background has a macroscopic correlation length. This modifies of the model pursued in \cite{utestat} on a long range scale.

\subsection{The dynamic generating functional}
A convenient mathematical tool to calculate dynamic correlation functions is the path integral formalism \cite{utestat,duesseldorfer}. We introduce the dynamic functional in Ito-discretization:
\ba
\lefteqn{{\cal Z}_M[{\{\bf h\}},{\{\tilde{\bf h}\}}] := 
\left\langle\; \e^{\tex\ i \int dt\sum_{i=1}^n 
\left( {\bf h}_i(t) {\bf r}_i(t) + {\tilde{\bf h}_i(t)} {\tilde{\bf r}_i(t)} \right)}\;\right\rangle}
\\
& &\hspace*{-.3cm}=\int d[{\bf r},\tilde{\bf r}] \;\e^{\tex - \int dt\sum_{i=1}^n
\left[ \gamma \tilde{\bf r}_i^2 
- i {\tilde{\bf r}_i}
\left( \dot{\bf r}_i + \gamma \frac{\dis\partial}{\dis\partial {\bf r}_i}{\cal H}_M\right) 
- i \left( {\bf h}_i {\bf r}_i + \tilde{\bf h}_i \tilde{\bf r}_i \right)
\right]}\:.\nonumber\\
\label{dynfun}
\ea
The time discretized meaning of the functional integral is $\int d[{\bf r},\tilde{\bf r}]=\int\prod_k\prod_j [d^dr_j(\tau_k)d^d\tilde r_j(\tau_k)/(2\pi)^d]$. All time integrals without limits are understood to run from $-\infty$ to $+\infty$. Differentiation with respect to the external fields $\{{\bf h}_i(t),{\tilde{\bf h}}_j(\tilde{t})\}$ yields correlation functions of $\{{\bf r}_i(t),{\tilde{\bf r}}_j(\tilde{t})\}$. 

As in \cite{utestat} we formally have to regularize the theory by adding to the Hamiltonian eq.~(\ref{ham}) a term
\be
{\cal H}_\Omega = \frac{{\bf R}_{cm}^2(t)}{2 L^2}\:,
\ee
where
\be
{\bf R}_{cm}(t) = \frac{1}{n} \sum_{i=1}^n {\bf r}_i(t)
\label{cms}
\ee
is the center of mass of the moving polymer. This term serves to locate the mobile chain in a finite volume of order $\Omega=L^d$. It is needed for purely technical reasons in intermediate steps of the calculation. We will let $L\to\infty$ as soon as possible. By means of the identity
\be
\frac{\partial}{\partial {\bf r}_l}\delta^d({\bf r}_l)= \int \frac{d^d p}{{(2\pi)}^d}\, i{\bf p}\, e^{i{\bf p}{\bf r}_l}
\label{deltaabl}
\ee
the dynamic functional can be written as
\be
{\cal Z}_M[\{{\bf h}\},\{\tilde{\bf h}\}] = 
\int d[{{\bf r},\tilde{\bf r}}] \;\e^{\tex - S_0 - S_I- S_I^{(B)} 
+ i \int dt\sum_{i=1}^n \left( {\bf h}_i {\bf r}_i + \tilde{\bf h}_i \tilde{\bf r}_i \right)}\:,
\label{dyfu}
\ee
where
\be
S_0 = \int dt \sum_{i=1}^n
\left[  \gamma  { \tilde{\bf r}}_{i}^2 
- i { \tilde{\bf r}}_{i}
\left( { \dot{\bf r}}_{i} + \gamma  \frac{2{\bf r}_{i} -
{ \bf r}_{i-1} - { \bf r}_{i+1}}{2l^2} + \gamma  \frac{{\bf R}_{cm}}{n L^2} \right)\right]
\label{rouse}
\ee 
is the free action and 
\be
S_I = u_0 l^d\int_{ \bf p} \int \gamma dt
\sum_{\stackrel{i,j=1}{j\neq i}}^n 
\left( { \tilde{\bf r}}_i(t){ \bf p}\right) 
\e^{\tex i { \bf p} ({\bf r}_{i}(t) - {\bf r}_j(t)) }
\label{inter}
\ee
arises from the excluded volume interaction among segments of the moving chain. The interaction with the background chains is contained in $S_I^{(B)}$:
\be 
S_I^{(B)} =  v_0 l^d\int_{ \bf p} \int d^d\hat{r}\,\e^{\tex -i{\bf p}\hat{\bf r}} \rho^{(B)}(\hat{\bf r})\int \gamma dt
\sum_{k=1}^n \left( { \tilde{\bf r}}_k(t){ \bf p}\right) 
\e^{\tex i {\bf p}{\bf r}_{k}(t) }\:.
\label{interb}
\ee
We here use the common abbreviation
\be
\int_{\bf p} \equiv \int \frac{d^d p}{{(2\pi)}^d}\:. 
\label{momin}
\ee
We now turn to the disorder average,
\be
\overline{{\cal Z}_M[\{{\bf h}\},\{\tilde{\bf h}\}]}=\int{\cal D}\{{\bf r}^{(B)}\} \,{\cal P}_M^{(B)}[\{{\bf r}^{(B)}\}]\:{{\cal Z}_M}\:. 
\label{Zdisav}
\ee 
Noting that the background coordinates enter ${\cal Z}_M$ only via $S_I^{(B)}$, we rewrite this average in terms of a cumulant expansion: Introducing 
\be 
\sigma(\hat{\bf r}) =  -v_0 l^d\int_{ \bf p} \int \gamma dt
\sum_{k=1}^n \left( { \tilde{\bf r}}_k(t){ \bf p}\right) 
\e^{\tex i {\bf p}({\bf r}_{k}(t)- \hat{\bf r})}
\label{abbr}
\ee
we obtain
\be
\overline{\e^{\tex - S_I^{(B)}}}=\overline{\e^{\tex\int d^d\hat{r}\sigma(\hat{\bf r})\rho^{(B)}(\hat{\bf r})}}=\e^{\tex\sum_{j=1}^\infty\frac{1}{j!}S^{(B)}_{I,j}}\:,
\label{quav}
\ee
where
\ba
S^{(B)}_{I,j}&=&\int d^d\hat{r}_1\ldots d^d\hat{r}_j\,
\sigma(\hat{\bf r}_1)\ldots\sigma(\hat{\bf r}_j)\,{\overline{\rho^{(B)}(\hat{\bf r}_1)\ldots\rho^{(B)}(\hat{\bf r}_j)}}^{\:\rm C}\nonumber\\
&=&{(-v_0 l^d)}^j\int\gamma dt_1\ldots\gamma dt_j\sum_{k_1,\ldots ,k_j=1}^{n}\int_{{\bf p}_1\ldots{\bf p}_j}\nonumber\\
& &\left({\tilde{\bf r}}_{k_1}(t_1){\bf p}_1\right)\ldots\left({\tilde{\bf r}}_{k_j}(t_j){\bf p}_j\right)\,\e^{\tex i {\bf p}_1{\bf r}_{k_1}(t_1)+\ldots + i {\bf p}_j{\bf r}_{k_j}(t_j)}\nonumber\\
& &{(2\pi)}^d{\delta}^{d}({\bf p}_1+\ldots + {\bf p}_j)\: I_{j}^{(B)}({\bf p}_1,\ldots, {\bf p}_{j-1})\:.
\label{quavac}
\ea
The static $j$-point density correlation function $I_{j}^{(B)}$ of the background polymers is defined as (note translational invariance):
\ba
\lefteqn{I_{j}^{(B)}({\bf p}_1,\ldots, {\bf p}_{j-1})=}
\nonumber\\& &\hspace*{-.2cm}
\int d^d\hat{r}_1\ldots d^d\hat{r}_{j-1}\e^{\tex -i{\bf p}_1{\hat{\bf r}}_{1}-\ldots -i{\bf p}_{j-1}{\hat{\bf r}_{j-1}}} \overline{\rho^{(B)}(\hat{\bf r}_1)\ldots\rho^{(B)}(\hat{\bf r}_{j-1})\rho^{(B)}(0)}^{\:\rm C}\:.\nonumber\\
\label{denscumu}
\ea
The superscript C stands for the cumulant. Now 
\be
I_{1}^{(B)}=\overline{\rho^{(B)}(0)}= c
\label{conc}
\ee
is the segment concentration of the background polymers. A short glance at eq.~(\ref{quavac}) reveals that $S^{(B)}_{I,1}\equiv 0$. Clearly a constant background potential cannot exert any force on the mobile chain. The next term $j=2$ in the sum eq.~(\ref{quav}) involves a single independent momentum integration, just as $S_I$, eq.~(\ref{inter}). Writing out these two contributions explicitly, we finally find the averaged dynamic generating functional as:
\ba
\lefteqn{\overline{{\cal Z}_M[{\{\bf h\}},{\{\tilde{\bf h}\}}]}=\int d[{\bf r},\tilde{\bf r}] \;\e^{\tex - S_0+i \int dt\sum_{i=1}^n 
\left( {\bf h}_i(t) {\bf r}_i(t) + {\tilde{\bf h}_i(t)} {\tilde{\bf r}_i(t)} \right)}}\nonumber\\
& &\!\exp\Bigg\{-\int_{\bf p}\bigg[u_0 l^d \int \gamma dt
\sum_{\stackrel{i,j=1}{j\neq i}}^n ( { \tilde{\bf r}}_i(t){ \bf p}) \e^{\tex i { \bf p} ({\bf r}_i(t) - {\bf r}_j(t))}+\nonumber\\
& &\!+\frac{1}{2} v_0^2 l^{2d} I_{2}^{(B)}(p) \int \gamma^2 \;dt_1
dt_2 \sum_{i,j=1}^n 
\Big({\tilde{\bf r}}_{i}(t_1){\bf p}\Big) \Big({\tilde{\bf r}}_{j}(t_2){\bf p}\Big)
\e^{\tex i {\bf p} ({\bf r}_{i}(t_1) - {\bf r}_j(t_2))}\bigg]\Bigg\} 
\nonumber\\
& &\!\exp{\Bigg\{\sum_{j=3}^\infty\frac{S^{(B)}_{I,j}}{j!}\Bigg\}}
\:.
\label{final}
\ea
The terms $S^{(B)}_{I,j}, j\ge3$, contribute only for higher orders (two loop) of perturbation theory and will not be explicitly considered here.
 
Before turning to an explicit calculation of the center of mass motion in the next section we should compare our general result with the one of \cite{utestat}. Ignoring $S^{(B)}_{I,j}, j\ge3$, we see that our result for the averaged generating functional resembles that in \cite{utestat}, except for the occurrence of the momentum-dependent density cumulant $I_{2}^{(B)}(p)$. The form of the latter is characteristic for the quenched background distribution. If we would have worked with a Gaussian distribution of unconnected beads, $\overline{\rho^{(B)}({\bf r})\rho^{(B)}({\bf r}')}^{\:\rm C}\sim \delta^d({\bf r}-{\bf r}')$, we would have found $I_{2}^{(B)}(p)\sim \mbox{const.}$, just as in \cite{utestat}. Formally we can map the theory as presented here to a theory with Gaussian random potentials by taking in the unrenormalized result eq.~(\ref{final}) the limit of infinite monomer density, $c\to\infty$. For $I_{2}^{(B)}(p)$ the limiting behaviour can be deduced from eq.~(\ref{Itree}), whereas a discussion of $j$-point density correlation functions can be found in \cite{Bal86}: 
\be
I_{j}^{(B)}({\bf p}_1,\ldots, {\bf p}_{j-1})\stackrel{c\to\infty}{=}\Bigg\{ \begin{array}{cc} 1/u_0^{(B)}l^d& \mbox{for}\quad j=2 \nonumber\\ 0 &\mbox{for}\quad j>2\end{array}\:.
\label{Iinf}
\ee
Thus, with $v_0^2/u_0^{(B)} =:v$, the generating functional eq.~(\ref{final}) becomes identical to that of \cite{utestat}. Perturbation theory then proceeds in the two couplings $u_0$ and $v$ only. This result has an appealing interpretation since the limit $c\to\infty$ lets one think of a dense polymer background, where the density correlation length should be microscopic indeed. However one should keep in mind that our model cannot describe dense systems properly, since it disregards three and higher body interactions etc. The limit $c\to\infty$ thus is purely formal.

\setcounter{equation}{0}
\section{Center of Mass Motion}
We aim at a renormalized one loop calculation of the center of mass motion,
\be
R^2(t):=\overline{\left\langle{\left({\bf R}_{cm}(t)-{\bf R}_{cm}(0)\right)}^2\right\rangle}\:.
\label{R2}
\ee
Remember that angular brackets denote the average over thermal forces, a bar stands for averaging over the quenched polymer background. In the formalism developed in sect.\ 2 we have
\be
R^2(t)=-{{\Delta}_{\bf q}}_{\vspace*{-0.1cm}{\big |}_0}\vspace*{+0.2cm}\overline{{\cal Z}_M[{\{{\bf h}_{cm}\}},{\{{\bf 0}\}}]}\:,
\label{R2Z}
\ee
where 
\be
{\bf h}_i(\tau)_{cm}=-\frac{\bf q}{n}\left( \delta(t-\tau)-\delta(-\tau)\right)
\label{hcm}
\ee
is independent of the segment index $i$. Expanding the generating functional eq.~(\ref{final}) in the couplings $u_0$ and $v_0$, we write:
\be
\overline{{\cal Z}_M[{\{{\bf h}_{cm}\}},{\{{\bf 0}\}}]}={\cal G}_{cm}(q,t)_0+{\cal G}_{cm}(q,t)_1+{\cal O}(2 \mbox{ loop}).
\label{zcm}
\ee
The tree approximation ${\cal G}_{cm}(q,t)_0$ can be found in \cite{utestat}, sect.\ 4.1. For system volume $L^d\to\infty$ (note that the system size $L$ is called $\xi$ in \cite{utestat}) we obtain:
\ba
{\cal G}_{cm}(q,t)_0&=&\int d[{\bf r},\tilde{\bf r}] \;\e^{\tex - S_0+i \int d\tau\sum_{i=1}^n 
\left( {\bf h}_i(\tau)_{cm} {\bf r}_i(\tau) \right)}\nonumber\\
&=& \e^{\tex -q^2 {{\Gamma}_0} t}\:,
\label{gtree}
\ea
where 
\be
{\Gamma_0}=\frac{\gamma}{n}
\label{Gamma}
\ee
is introduced. Considering the one loop contribution we note that the center of mass motion of a Rouse chain with excluded volume (= our model without disorder, i.e. $v_0=0$) is independent of the excluded volume interaction. Therefore all terms in an expansion of $\overline{{\cal Z}_M[{\{{\bf h}_{cm}\}},{\{0\}}]}$, which are proportional to a power of $u_0$ only, strictly vanish. We thus find:
\ba
{\cal G}_{cm}(q,t)_1=-\frac{1}{2} v_0^2 l^{2d}\int d[{\bf r},\tilde{\bf r}] \;\e^{\tex - S_0+i \int d\tau\sum_{i=1}^n 
\left( {\bf h}_i(\tau)_{cm} {\bf r}_i(\tau) \right)}\qquad& &\nonumber\\
\int_{\bf p} I_{2}^{(B)}(p) \int \gamma^2 \;dt_1
dt_2 \sum_{i,j=1}^n 
\Big({\tilde{\bf r}}_{i}(t_1){\bf p}\Big) \Big({\tilde{\bf r}}_{j}(t_2){\bf p}\Big)
\e^{\tex i {\bf p} ({\bf r}_{i}(t_1) - {\bf r}_j(t_2))}& &
\label{gloop}
\ea
Except for the integration over $\bf p$, which involves the momentum dependent density correlation function $I_{2}^{(B)}(p)$, we now have to carry through the same calculational steps as in \cite{utestat}, sect.\ 4.2. Taking over the interim result before performing the momentum integration from there yields: 
\ba
\overline{{\cal Z}_M[{\{{\bf h}_{cm}\}},{\{{\bf 0}\}}]}\!\!&=&\!\!\e^{\tex -q^2 {\Gamma_0} t}\Bigg\{1+v_0^2l^{2d}{\Gamma_0} t\int_{\bf p} I_{2}^{(B)}(p) ({\bf pq})^2\int_0^t {\Gamma_0}\, d\tau\,(1-\tau/t)\nonumber\\
& & \sum_{i,j=1}^n \e^{-\tex p^2 D_{ij}(\tau)+2{\bf pq}{\Gamma_0}\tau} +{\cal O}(2 \mbox{ loop})\Bigg\}\:.
\label{interim}
\ea
Here the limit $L\to\infty$ has been carried through. $D_{ij}(t)$ denotes a segment-segment correlation function of the Rouse model:
\be
D_{ij}(t):=\frac{1}{2d}\left\langle{\left({\bf r}_{i}(t)-{\bf r}_{j}(0)\right)}^2\right\rangle_0=\frac{1}{2d}\int d[{\bf r},\tilde{\bf r}] \;{\left({\bf r}_{i}(t)-{\bf r}_{j}(0)\right)}^2\;\e^{\tex - S_0}\:.
\label{Ddef}
\ee
An extensive discussion of this quantity can be found in \cite{utestat}, appendix A. 

As a next step we take the momentum derivative in eq.~(\ref{R2Z}) and introduce variables $y={\bf p}_{\scriptstyle\perp}^2, z=p_{\scriptscriptstyle\parallel}^2$, where the perpendicular or parallel direction are defined with respect to $\bf q$. Thus $p^2=y+z$ and ${\bf pq}=\pm\sqrt{zq^2}$. After substituting $z$ by $\hat{z}=D_{ij}(\tau)(y+z)$,  
we perform the $y$-integration and obtain:  
\ba
R^2(t)&=&2d{\Gamma_0} t\,\Bigg\{1-v_0^2l^{2d}\frac{(4\pi)^{-d/2}}{2{\Gamma}(1+d/2)}\int_0^t \,{\Gamma_0} d\tau \left(1-\tau/t\right)\sum_{i,j=1}^n {D_{ij}(\tau)}^{-1-d/2} \nonumber\\
& &\qquad\qquad\quad\int_0^\infty d\hat{z}\,{\hat{z}}^{d/2} e^{-\hat{z}} I_{2}^{(B)}(\sqrt{\hat{z}/D_{ij}(\tau)}) +{\cal O}(2\mbox{loop})\Bigg\}\:.\nonumber\\
\label{Rzwifi}
\ea

We now insert the tree approximation for the two point correlation function of the background which takes a form well known from random phase approximation \cite{DGeJaRPA}:
\be
I_{2}^{(B)}(q)= \frac{cND_p(Nq^2l^2)}
{1+u_{0}^{(B)}l^dcND_p(Nq^2l^2)}+{\cal O}(1 \mbox{ loop})\:.
\label{Itree}
\ee
The Debye function averaged over polydispersity reads 
\be
D_p(x)= \frac{2}{x^2}\left(\tilde{p}(x)-1+x\right)\:.
\label{Debpol}
\ee
The function $\tilde{p}(x)$ introduced here is the Laplace transform of the reduced chain length distribution $p(y):= Nc_p(yN)/c_p$ \cite{SchWi80}. For a monodisperse ensemble, $p(y)=\delta(y-1)$, we have $\tilde{p}(x)=e^{-x}$, which in eq.~(\ref{Debpol}) yields the standard Debye function. An expansion of $\tilde{p}(x)$ for small $x$ reads $\tilde{p}(x)=1-x+\tilde{p}''x^2/2+\ldots$, such that $D_p(0)=\tilde{p}''$ exists.  

In our result eq.~(\ref{Rzwifi}) we now take the continuous chain limit $l\to 0$, $n\to\infty$ with $S:=nl^2=\mbox{const}$. This amounts to replacing the sums over segments by integrals: $l^{4}\sum_{i,j=1}^{n}=\int_0^Sds_ids_j$. We then can make all quantities dimensionless by dividing by the appropriate power of $S$. Denoting $\epsilon=4-d$, our final unrenormalized one loop result reads:
\ba
R^2(t)&=&2d{\Gamma_0} t\:\Bigg\{1-\frac{v_0 n^{\epsilon/2}}{2{(4\pi)}^{d/2}{\Gamma}(1+d/2)} \int_0^t \frac{{\Gamma_0} d\tau}{S}\, \left(1-\tau/t\right)\nonumber\\
&&\hspace{2.0cm}\int_0^S\frac{ds_ids_j}{S^2}\,{\left(\frac{D_{ij}(\tau)}{S}\right)}^{-1-d/2}\int_0^\infty d\hat{z}\,{\hat{z}}^{d/2} e^{-\hat{z}}\nonumber\\
&&\hspace{2.0cm}\frac{v_0l^dcND_p\left(Nl^2\hat{z}/D_{ij}(\tau)\right)}{1+u_0^{(B)}l^dcN\,D_p\left(Nl^2\hat{z}/D_{ij}(\tau)\right)}+{\cal O}(2\mbox{loop})\Bigg\}\:.\nonumber\\
\label{Rzwisec}
\ea

\setcounter{equation}{0}
\section{Renormalization}
Eq.~(\ref{Rzwisec}) exhibits a general feature of perturbation expansions in polymer theory: The expansion in coupling constants in fact is an expansion in (coupling constant) $\cdot$ $(\mbox{segment number})^{\epsilon/2}$. This can also be seen in the dynamic generating functional by general dimensional arguments (power counting) as explained in \cite{utestat}, sect.\ 3.4. For long chains bare perturbation theory in dimension $d<4$ thus breaks down. Now the solution to this problem is well known. One can map the polymers with microscopic segment size $l$ on equivalent ones with greater effective segments of renormalized size $l_R$. Since the total extension of the polymers as measured e.g.\ by the radius of gyration remains constant, the renormalized segment number must decrease. This way we obtain a renormalized expansion parameter (ren.\ coupling constant) $\cdot$ $(\mbox{ren.\ segment number})^{\epsilon/2}$, which stays finite even for long chains. 

The construction of such a renormalization group (RG) mapping amounts to extracting the dependence of the theory on a change of the microscopic scale $l$ and to absorb it into the RG mapping. In the continuous chain formulation $l\to 0$ this microstructure dependence is hidden in divergences ($\epsilon$-poles) which occur, when the space dimension reaches the upper critical dimension $d=4$. In such a dimensionally regularized theory the RG mapping is set up by introducing renormalization factors which absorb the $\epsilon$-poles of the bare theory. To fix these renormalization factors and thus the RG mapping, one has to calculate appropriate observables up to the desired loop order. This task here however is greatly simplified by the observation that polymer statics is contained in the dynamic theory in the sense that one equally could calculate all static quantities in the framework of the full dynamic theory. This means that the renormalization of all parameters that occur also in polymer statics, i.e.\ all except for ${\Gamma_0}$, can be found by an inspection of the static theory. This is done in the next subsection.

\subsection{Static renormalization: Quenched vs.\ annealed}
Calculating a static observable ${\cal O}_S$ like the radius of gyration, we do not need the full dynamic theory, but we can employ the static equilibrium distribution instead:
\be
{\cal P}_{eq.}[\{{\bf r}\};\{{\bf r}^{(B)}\}]=\frac{1}{{\cal Z}_M[\{{\bf r}^{(B)}\}]}\: \e^{\tex -{\cal H}_M[\{{\bf r}\};\{{\bf r}^{(B)}\}]}\:. 
\label{stateq}
\ee
Note that ${\cal P}_{eq.}$ is taken for a single realization of the background distribution and thus the $\{{\bf r}^{(B)}\}$ are to be considered as parameters, not as proper degrees of freedom. Note further that in this equilibrium model the segment coordinates $\{{\bf r}\}$ of the test chain are independent of $t$. The partition function of the test chain takes the form
\be
{\cal Z}_M[\{{\bf r}^{(B)}\}]=\int{\cal D}\{{\bf r}\}\, \e^{\tex -{\cal H}_M[\{{\bf r}\};\{{\bf r}^{(B)}\}]}\:, 
\label{Zque}
\ee
where
\be
{\cal D}\{{\bf r}\}= \prod_{i=1}^{n}\frac{d^dr_{i}}{(4 \pi l^2)^{d/2}}\:. 
\label{meassing}
\ee
The Hamiltonian ${\cal H}_M$ is given in eq.~(\ref{ham}), with ${\bf r}_i(t) \to{\bf r}_i$. The static average of ${\cal O}_S$ then reads
\be
\overline{\left\langle{{\cal O}_S}\right\rangle}=\overline{\int{\cal D}\{{\bf r}\} \,{\cal P}_{eq.}[\{{\bf r}\};\{{\bf r}^{(B)}\}]\,{{\cal O}_S}[\{{\bf r}\};\{{\bf r}^{(B)}\}]} 
\label{statobs}
\ee
and exhibits the well known problem of quenched static averages, namely the occurrence of the disorder dependent partition function ${\cal Z}_M$ in the denominator of eq.~(\ref{stateq}), which in general makes static averages over quenched disorder (i.e.\ the bar in eq.~(\ref{statobs})) difficult to perform. Now the well known solution to this problem is the observation, that the partition function is self-averaging, as long as the system size $\Omega$ is infinitely large compared to any other relevant length scale \cite{CatesBall,WuHui,utemacro}. This argument also applies here, since we take the thermodynamic limit $\Omega\to\infty$ with keeping chain lengths and hence the background correlation length $\xi^{(B)}$ as well as the radii of gyration, $R_g$ resp.\ $R_g^{(B)}$ fixed and finite. We rephrase the self-averaging argument in appendix A. Let us now discuss the consequences:

By means of the self-averaging property, ${\cal Z}_M=\overline{{\cal Z}_M}$, we can write eq.~(\ref{statobs}) as:
\be
\overline{\left\langle{{\cal O}_S}\right\rangle}=\overline{\int{\cal D}\{{\bf r}\} \,{{\cal O}_S}[\{{\bf r}\};\{{\bf r}^{(B)}\}]\,\e^{\tex -{\cal H}_M[\{{\bf r}\};\{{\bf r}^{(B)}\}]}}{\Big /}\overline{{\cal Z}_M[\{{\bf r}^{(B)}\}]}\:. 
\label{statobsan}
\ee
Now we have a separate average in numerator and denominator, i.e.\ an average as in the corresponding annealed system. But such a system is nothing else than the well known static ternary system of two polymer species in solution \cite{Joa84,SchLeKa91,SchKa93}. Indeed, taking into account eq.~(\ref{cano}) we can write out the average in eq.~(\ref{statobsan}) as:   
\be
\overline{\left\langle{{\cal O}_S}\right\rangle}=\frac{1}{{\cal Z}}\int{\cal D}\{{\bf r},{\bf r}^{(B)}\} \,{{\cal O}_S}[\{{\bf r}\};\{{\bf r}^{(B)}\}]\,\e^{\tex -{\cal H}_M[\{{\bf r}\};\{{\bf r}^{(B)}\}]-{\cal H}_M^{(B)}[\{{\bf r}^{(B)}\}]}\:. 
\label{expl}
\ee
As usual the partition function ${\cal Z}$ denotes the integral in eq.~(\ref{expl}) with ${\cal O}_S\equiv 1$. This is to be compared with static averages in annealed ternary systems. The chains in a ternary solution of $M_1$ polymers of sort 1 and $M_2$ polymers of sort 2 interact via three couplings $u_0^{(ab)}=u_0^{(ba)}; a,b\in\{1,2\}$; representing the intrachain $(a=b)$ resp.\ interchain $(a\neq b)$ repulsion. Together with the chain connectedness the Hamiltonian then reads:
\ba
\lefteqn{{\cal H}_{12}[\{{\bf r}^{(1)};{\bf r}^{(2)}\}] =} 
\nonumber\\& &
\hspace{-.1cm}\frac{1}{4l^2}\sum_{a=1}^{2}\sum_{m=1}^{M_a}\sum_{j=2}^{n_m^{(a)}}({\bf r}_{m,j}^{(a)} -
{\bf r}_{m,j-1}^{(a)})^2 + \frac{1}{2}\sum_{a,b=1}^{2}u_0^{(ab)}l^d\int d^dr\,\rho^{(a)}({\bf r})\rho^{(b)}({\bf r})\:,\nonumber\\
\label{terham}
\ea
where
\be
\rho^{(a)}({\bf r})= \sum_{m=1}^{M_a}\sum_{j=1}^{n_m^{(a)}}\delta^d({\bf r}-{\bf r}_{m,j}^{(a)})
\label{denster}
\ee
is the segment density of the polymer species $a=1,2$. Let the tracer chain in our original system be of sort 1, i.e.\ $M_1=1$, $n_1^{(1)}=n$, the background chains be of sort 2. Comparing ${\cal H}_M+{\cal H}_M^{(B)}$, eq.~(\ref{expl}), with the ternary Hamiltonian as above, we indeed find complete agreement. The couplings can be identified as the three ternary couplings:
\be
u_0=u_0^{(11)}, u_0^{(B)}=u_0^{(22)}, v_0=u_0^{(12)}\:.
\label{coupall}
\ee
Having established the connection of the static behaviour of our model to annealed ternary polymer systems, we now can simply adopt the corresponding renormalization scheme. The one loop results we quote here can be found in \cite{SchKa93} (note that we set the parameter $b_u$ introduced there to $b_u\equiv 1$ from the outset). 
The mapping of the bare static parameters $(l; n, N; u_0, u_0^{(B)}, v_0)$ to the corresponding renormalized quantities $(l_R; n_R, N_R; u, u^{(B)}, v)$ takes the form:
\ba
nl^2&=&n_R l_R^2 \left\{1-\frac{u}{\epsilon}+{\cal O}(2\mbox{loop})\right\}\label{nrenor}\\
Nl^2&=&N_R l_R^2 \left\{1-\frac{u^{(B)}}{\epsilon}+{\cal O}(2\mbox{loop})\right\}
\label{Nrenor}\\
(4 \pi)^{-d/2}u_0l^{-\epsilon}&=&ul_R^{-\epsilon}\frac{1}{2}
\left\{1+\frac{4u}{\epsilon} +{\cal O}(2\mbox{loop})\right\}
\label{urenor}\\
(4 \pi)^{-d/2}u_0^{(B)}l^{-\epsilon}&=&u^{(B)}l_R^{-\epsilon}\frac{1}{2} \left\{1+\frac{4u^{(B)}}{\epsilon}+{\cal O}(2\mbox{loop})\right\}
\label{ubrenor}\\
(4 \pi)^{-d/2}v_0l^{-\epsilon}&=&vl_R^{-\epsilon}\frac{1}{2} \left\{1+\frac{u+u^{(B)}}{\epsilon}+\frac{2v}{\epsilon}+{\cal O}(2\mbox{loop})\right\}
\label{vrenor}
\ea
The renormalization of $l, n, u_0$ resp.\ $l, N, u_0^{(B)}$ is just as in the corresponding binary systems. Only the ternary parameter $v_0$ is renormalized by all three couplings.

Now aiming at a renormalization of the center of mass motion a glance at eq.~(\ref{Rzwisec}) immediately reveals that any possible divergence in the one loop integral could be subtracted only by a renormalization of ${\Gamma_0}$, since this is the only parameter that occurs already in the tree result. All $\epsilon$-poles of eqs.~(\ref{nrenor})-(\ref{vrenor}) contribute, if inserted into eq.~(\ref{Rzwisec}), only in 2 loop. This was to be expected, since $R^2(t)$ is a purely dynamic quantity. The center of mass motion in one loop thus fixes the one loop renormalization of ${\Gamma_0}$ only, to which we will turn now.

\subsection{Dynamic renormalization}

We now search for possible $\epsilon$-poles in the one loop contribution to eq.~(\ref{Rzwisec}). Divergences are to be expected for large frequencies and momenta (UV-divergences), that is to say for short times and short segment separation. We therefore introduce appropriate dimensionless time and distance variables:
$\hat{\tau}:={\Gamma_0}\tau/S$ and ${\hat y}:=(s_i-s_j)/S/\sqrt{2\hat{\tau}}$ together with ${\hat s}:=s_j/S$. In these variables, the segment-segment correlation function $D_{ij}(\tau)$, eq.~(\ref{Ddef}), can be written as:
\be
\frac{D_{ij}(\tau)}{S}=\sqrt{2\hat{\tau}}\,F({\hat y},{\hat y}+2{\hat s}/\sqrt{ 2\hat{\tau}},2/\sqrt{2\hat{\tau}})\:.
\label{F}
\ee
The properties of the function $F(y,z,\lambda)$ are discussed in great detail in appendix A of \cite{utestat}. We here simply cite the following two representations, the latter being suitable for an analysis of the behaviour for short times and short segment separation:
\ba
F(y,z,\lambda)&=&\frac{1}{\lambda}+\lambda\left(\frac{1}{3}-\frac{z}{\lambda}+\frac{y^2+z^2}{{\lambda}^2}\right)\nonumber\\
& &-\sum_{k=1}^{\infty}\frac{\lambda}{{\pi}^2k^2} e^{-{\pi}^2k^2/{\lambda}^2}\left(\cos \frac{2{\pi}ky}{\lambda}+\cos \frac{2{\pi}kz}{\lambda}\right)
\label{Fgr}\\
&=&|y|+\sum_{\nu=-\infty}^{\infty}\left[f(y+\nu\lambda)-|y+\nu\lambda|+f(z+\nu\lambda)-|z+\nu\lambda|\right]\:.\nonumber\\
\label{Fkl}
\ea
The function $f(y)$ is defined as
\be
f(y)=\frac{e^{-y^2}}{\sqrt{\pi}} + y\,{\rm erf}(y)\:,
\label{f}
\ee
where ${\rm erf}(y)$ denotes the error function. The one loop result eq.~(\ref{Rzwisec}), written completely in dimensionless variables then reads:
\ba
\frac{R^2(t)}{2d{\Gamma_0} t}&=&\hspace{-0.0cm}1-\frac{v_0 n^{\epsilon/2}}{2(32{\pi}^2)^{d/4}{\Gamma}(1+d/2)}\int_0^T d\hat{\tau}\, {\hat{\tau}}^{-d/4} \left(1-\hat{\tau}/T\right)\int_0^\infty d\hat{z}\,{\hat{z}}^{d/2}e^{-{\hat{z}}}\nonumber\\
&&\hspace{+0.00cm} \int_0^1 d{\hat{s}}\,\int_{-{\hat{s}}/\sqrt{2\hat{\tau}}} ^{(1-{\hat{s}})/\sqrt{2\hat{\tau}}} d{\hat{y}}\,F({\hat y},{\hat y}+2{\hat s}/\sqrt{ 2\hat{\tau}},2/\sqrt{2\hat{\tau}})^{-1-d/2} H({\hat{\tau}},{\hat{z}},{\hat{s}},{\hat{y}})\nonumber
\ea 
\vspace*{-.365cm}
\be 
\hspace*{-7.2cm}+{\cal O}(2\mbox{loop})\:.
\label{Rendu}
\ee
The dimensionless time variable $T$ is
\be
T=\frac{{\Gamma_0} t}{n l^2}\:.
\label{T}
\ee
The function $H$ has its origin in the tree approximation of the density correlation function:
\be
H({\hat{\tau}},{\hat{z}},{\hat{s}},{\hat{y}})= \frac{v_0l^dcN D_p\left(\frac{ N}{ n}\frac{ {\hat{z}}}{ \sqrt{2{\hat{\tau}}}F({\hat y},{\hat y}+2{\hat s}/\sqrt{ 2\hat{\tau}},2/\sqrt{2\hat{\tau}})}\right)}
{ 1+u_0^{(B)}l^dcN\,D_p\left(\frac{ N}{ n}\frac{ {\hat{z}}}{ \sqrt{2{\hat{\tau}}}F({\hat y},{\hat y}+2{\hat s}/\sqrt{2\hat{\tau}},2/\sqrt{2\hat{\tau}})  }\right)}\:.
\label{H}
\ee
In the following we will drop the hats in the integration variables $\hat{\tau},\hat{z},\hat{s}$ and $\hat{y}$.

Let us first discuss what happens in the case of \cite{utestat}. To this end we as before take the formal limit $c\to\infty$ of infinite segment concentration, such that $H\to v_0/u_0^{(B)}$. The divergent part in $d=4$ then comes from the integration:
\be
\int_0^T d{\tau}\, {{\tau}}^{-d/4} \int_0^1 d{{s}}\,\int_{-{{s}}/\sqrt{2{\tau}}} ^{(1-{{s}})/\sqrt{2{\tau}}} d{{y}}\,F({ y},{ y}+2{ s}/\sqrt{ 2{\tau}},2/\sqrt{2{\tau}})^{-1-d/2}
\label{diver}
\ee
For small $\tau$ we can replace the right hand integrations by
\be
\int_0^1 ds\int_{-\infty}^{\infty}dy\,F\left(y,\infty,\infty\right)^{-1-d/2},
\label{diver1}
\ee
where $F(y,\infty,\infty)=f(y)$, which can easily be derived from eq.~(\ref{Fkl}), taking into account the asymptotic behaviour $f(x)\stackrel{x\to\pm\infty}{\sim}|x|$. In four dimensions eq.~(\ref{diver1}) thus yields two times the finite constant
\be
I:=\int_0^\infty dy\;f(y)^{-3} \;=\; 3.587 \ldots\:  .
\label{I}
\ee
The remaining $\tau$-integral diverges for $d=4$ at the lower bound:
\be
\int_0^T d\tau\, {\tau}^{-d/4}=\frac{4}{\epsilon}\left(1+{\cal O}(\epsilon)\right)\:.
\label{pole}
\ee
All together this ends up in a one loop divergence $\sim I/\epsilon$, which has to be absorbed in a renormalization of ${\Gamma_0}$.

Now in our system, where $c$ is small and finite, the long ranged density correlation function renders the one loop term finite in four dimensions. This is not hard to prove if one takes into account that the Debye function obeys (cf.\ eq.~(\ref{Debpol}))
\be
D_p(x) \stackrel{x\to\infty}{\sim}2/x\:,
\label{Debasy}
\ee
where the right hand side actually is an upper bound. This means that for $H$, eq.~(\ref{H}), the following inequality holds:
\be
0<H(\tau,z,s,y)<\frac{2v_0l^dcn}{z}\sqrt{2\tau}F(y,y+2s/\sqrt{ 2\tau},2/\sqrt{2\tau})\:.
\label{Hineq}
\ee
Hence an additional factor $\sqrt{\tau}$ occurs, which alters the power of $\tau$ in eq.~(\ref{Rendu}) from ${\tau}^{-d/4}$ to ${\tau}^{1/2-d/4}$ and thus improves the convergence for small values of $\tau$. Together with $F(y,z,\lambda)> F(y,\infty,\infty)$ it is easy to show that the right hand side of eq.~(\ref{Rendu}) exists for $2<d<6$. To see the role of a finite correlation length ${\xi^{(B)}}$ in this connection more clearly we note that the unrenormalized tree approximation of the latter in the semidilute limit reads (cf.\ eq.~(\ref{Itree})):
\be
{\xi^{(B)}}^2= -\frac{1}{2}{{\Delta}_{\bf q}}_{\vspace*{-0.1cm}{\big |}_0}\vspace*{+0.2cm}\ln I_{2}^{(B)}(q)\sim \frac{R_g^2}{u_0^{(B)}l^dcn}\:.
\label{xid}
\ee
Concentrating on the dangerous short time part of the time integral in eq.~(\ref{Rendu}) we according to eqs.~(\ref{H}),(\ref{Debasy}) write:
\ba
\lefteqn{\int_0 d{\tau}\, {{\tau}}^{-d/4}\frac{2v_0l^dcn\sqrt{2{\tau}}f(y)/z}{1+2u_0^{(B)}l^dcn\sqrt{2{\tau}}f(y)/z}=}\nonumber\\& &
\frac{v_0}{u_0^{(B)}}\int_0 d{\tau}\,{\tau}^{1/2-d/4}{\left[{\sqrt{\tau}+
\frac{ z}{ 2\sqrt{2}f(y)}{\left({\frac{ \xi^{(B)}}{ R_g}}\right)}^2}\,\right]}^{-1}\:.
\label{cutoff}
\ea
In case of a microscopic correlation length ${\xi^{(B)}}\to 0$ the time integral in $d=4$ would not exist.

The essence of this analysis is that up to the order of one loop ${\Gamma_0}$ needs no renormalization:
\be
{\Gamma_0}=\Gamma_R \left\{1+{\cal O}(2\mbox{loop})\right\}\:.
\label{GammaR}
\ee

\subsection{Renormalized center of mass motion}
Now we are prepared to renormalize the one loop result eq.~(\ref{Rendu}). Since in this loop order no $\epsilon$-poles show up, this procedure amounts to a simple replacement of the bare parameters by their renormalized counterparts according to eqs.~(\ref{nrenor})-(\ref{vrenor}), (\ref{GammaR}). Note that $c_p=c/N$ is not renormalized, since neither the volume $\Omega$ nor the number of polymer chains $M$ in the background is altered by renormalization. The reduced chain length distribution function (see eq.~(\ref{Debpol})) is a normalized function of the RG invariant ratio $n^{(B)}/N$ and thus is invariant under renormalization itself. Denoting the renormalized time as
\be
T_R=\frac{\Gamma_R t}{n_R l_R^2}\:,
\label{TR}
\ee
we obtain:
\ba
\frac{R^2(t)}{2dn_Rl_R^2T_R}\!&=&\!1-\frac{v n_R^{\epsilon/2}}{2^{2+d/4}{\Gamma}(1+d/2)}\int_0^{T_R} d\tau\, {\tau}^{-d/4} \left(1-\tau/T_R\right)\int_0^\infty dz\,z^{d/2}e^{-z}
\nonumber\\&& 
\quad\:\:\:\int_0^1 d{ {s}}\,\int_{-{ {s}}/\sqrt{2 {\tau}}} ^{(1-{ {s}})/\sqrt{2 {\tau}}} d{ {y}}\,F({  y},{  y}+2{  s}/\sqrt{ 2 {\tau}},2/\sqrt{2 {\tau}})^{-1-d/2}\nonumber
\ea 
\vspace*{-.3cm}
\be 
\hspace*{-1.9cm}H_R(\tau,z,s,y)+{\cal O}(2\mbox{loop}) \:.
\label{Rresren}
\ee
The renormalized function $H_R$ reads
\be
\hspace*{-.3cm}H_R(\tau,z,s,y)
=\frac{vc_RN_RD_p\Big({\tex\frac{N_R}{n_R}\frac{z}{\sqrt{2\tau}F(y,y+2s/\sqrt{2\tau},2/\sqrt{2\tau})}}\Big)}
{1+u^{(B)}c_RN_RD_p\Big({\tex\frac{N_R}{n_R}\frac{z}{\sqrt{2\tau}F(y,y+2  s/\sqrt{2\tau},2/\sqrt{2\tau})}}\Big)}\:,
\label{HR}
\ee
where we defined a dimensionless renormalized concentration as 
\be
c_R=\frac{{(4\pi)}^{d/2}}{2}l_R^dc_pN_R\:.
\label{cR}
\ee
To make use of a result expressed in renormalized parameters we have to find the explicit RG-mapping, which connects them with the physical parameters of the bare theory. This is a standard procedure of renormalized perturbation theory, which is done by an integration of flow equations. The latter are obtained by taking the partial derivative $-l_R\partial /\partial l_R$ of eqs.~(\ref{nrenor})-(\ref{vrenor}), keeping all bare parameters fixed. Let us briefly rephrase some well known results concerning the flow of the couplings in ternary polymer systems \cite{Joa84,SchLeKa91,SchKa93}: The RG flow of the three couplings has 8 fixed points. In the limit of long chains, $l_R\to\infty$, there is only one globally stable (IR-stable) fixed point, attracting the flow from all directions. This is the symmetric excluded volume fixed point, where all three couplings attain the same fixed point value:
\be
u^\ast={u^{(B)}}^\ast=v^\ast=\frac{1}{4}\epsilon+\frac{21}{128}\epsilon^2+{\cal O}(\epsilon^3)\:.
\label{symfi}
\ee
At this fixed point the RG mapping takes the simple form \cite{SchKa93}:
\be
l_R=B_1{\left(\frac{n}{n_R}\right)}^\nu=B_2{\left(\frac{N}{N_R}\right)}^\nu\:,      
\label{map}
\ee
where $B_1, B_2$ denote nonuniversal constants, which depend on the chemical microstructure of the two polymer species. The critical exponent $\nu$ governs the size of an isolated excluded volume chain without disorder as measured e.g.\ by the radius of gyration, $R_g\sim n^\nu$:
\be
\nu=\frac{1}{2}+\frac{1}{16}\epsilon+{\cal O}(\epsilon^2)
\label{nu}
\ee

All the other fixed points are repulsive in at least one direction. The symmetric Gaussian fixed point $u=u^{(B)}=v=0$, in particular, is globally repulsive. To reach such fixed points it needs a careful adjustment of the nonuniversal parameters of the system. In the sequel we therefore restrict ourselves to the symmetric excluded volume fixed point eq.~(\ref{symfi}), which governs the generic behaviour in the limit of long chains. A brief discussion of other fixed points can be found in appendix B. We note that we could evaluate our theory also in all the crossover regime among the different fixed points. Due to the lack of data we however do not pursue this program.

We should stress that the existence of stable fixed points is necessary for the occurrence of pure power laws, but by no means necessary for an application of RG methods. In the case of Gaussian disorder \cite{utestat} there is in fact no such fixed point in the coupling, which leads to nonlinear scaling behaviour. As we will see in the next section, the occurrence of a stable fixed point here in the semidilute limit leads to asymptotic dynamic power laws with new exponents.  

Having fixed the RG mapping one final step is to be done, namely to choose the renormalized length scale. The expansion parameter $v_0n^{\epsilon/2}\stackrel {\mbox{\scriptsize ren.}}{\to}vn_R^{\epsilon/2}$ in our renormalized result eq.~(\ref{Rresren}) suggests to rely on the standard choice $n_R\sim {\cal O}(1)$, which guarantees a finite expansion parameter. This choice implicitly fixes the renormalized length scale $l_R$ to be of the order of the extension of the moving coil, since in tree approximation the radius of gyration behaves as $R_g^2\sim nl^2\stackrel{\mbox{\scriptsize ren.}}{\to}n_Rl_R^2$. We should mention that although this choice is always possible, in many cases it is not the best one. It, for instance, in the semidilute limit leads to logarithmic singularities, which have to be cured by reexponentiating the $\epsilon$-expansion. In favorite cases such problems can be avoided by a more sophisticated choice of $l_R$. Quite generally one can say that the best and most physical choice of the renormalized length scale is to choose it of the order of the smallest relevant length scale in the problem. For the static problem of a single polymer in solution there is only one macroscopic length scale $R_g$, thus $l_R\sim R_g$. In case of semidilute polymer solutions however the blob size $\xi_c$ (concentration blobs) is the appropriate smallest macroscopic length scale \cite{Sch84}, while in the context of polymer dynamics time blobs $\xi_t$ have been invented successfully \cite{EbBaSch}. Trying to generalize these ideas to our system, where both $\xi_c$ and $\xi_t$ show up, it turns out that they lead to divergences in the one loop result which can not be handled. This could have been expected. The reason is that any blob idea is based on the existence of two separated regions: Inside the blob, where interactions are present and outside the blob, where we essentially have a "free" theory without nontrivial interactions or correlations. It is this separation that does not work in our case: Looking at short times, such that the time blob is the smallest length scale, $\xi_t<\xi_c$, we have inside of the blob dynamic correlations, but on scales larger than $\xi_t$ the theory is not trivial since the segments still can be inside a concentration blob, yielding nontrivial correlations. In the Gaussian case \cite{utestat,EbBaSch} this problem does not arise since the correlation length of the disorder is microscopic: $\xi_c\sim l\to 0$. For longer times such that $\xi_c<\xi_t$ we run into the same problems: Although the interaction is screened on scales $>\xi_c$, we still have nontrivial dynamic correlations. Aiming at a calculation of dynamic quantities these correlations ruin the simple picture of a chain being composed of uncorrelated concentration blobs as in the static theory of semidilute polymer systems.

We therefore rely on the $\epsilon$-expansion together with the standard choice of the renormalized length scale by fixing $n_R=1$. Furthermore we set the couplings to their fixed point value $u^{\ast}$. Eq.~(\ref{map}) then fixes all renormalized quantities that show up on the right hand side of eq.~(\ref{Rresren}) as a function of the bare parameters $n,N,c,$ and $t$. The bare parameters themselves then occur in certain dimensionless scaling combinations only. The corresponding scaling variables $\bar{Y},\bar{S},$ and $\bar{T}$ are defined as follows:
\ba
N_R&\stackrel{n_R=1}{=}&\frac{N_R}{n_R}=\frac{N}{n}{\left(\frac{B_2}{B_1}\right)}^{1/\nu}=:\bar{Y}
\label{yquer}\\
{u}^\ast c_R&=&\frac{{(4{\pi})}^{d/2}}{2}\,{u}^\ast B_2^d c_pN^{d\nu}{N_R}^{1-d\nu}
\nonumber\\
&\stackrel{n_R=1}{=}&cn^{d\nu-1}\frac{{(4{\pi})}^{d/2}}{2}\,{u}^\ast{\left(\frac{B_2}{B_1}\right)}^{1/\nu}B_1^d=:\bar{S}
\label{s}\\
T_R&\stackrel{n_R=1}{=}&\frac{\Gamma_R t}{l_R^2}=\frac{t}{n^{1+2\nu}}\frac{\gamma}{B_1^2}=:\bar{T}\:.
\label{time}
\ea
Regarding $T_R$ one should remember that up to the order of one loop ${\Gamma_0}$ is not renormalized, i.e.\ $\Gamma_R={\Gamma_0}=\gamma/n$, cf.\ eqs.~(\ref{Gamma}),(\ref{GammaR}).

\setcounter{equation}{0}
\section{Results}
Having finished the previous, more technical section, we here first collect and interpret our general results for the center of mass motion: The quantity we want to investigate is the normalized center of mass motion
\be
{\cal F}:=\frac{R^2(t)}{2dn_Rl_R^2}\:.
\label{Ferg}
\ee
It describes the mean squared center of mass displacement $R^2(t)$ of a chain of length $n$, divided by the radius of gyration $R_g^2\sim n_Rl_R^2 \sim B_1^2 n^{2\nu}$ of the tracer chain without disorder. According to the discussion at the end of the last section this function depends on three dimensionless scaling variables as defined in eqs.~(\ref{yquer})-(\ref{time}):
\be
{\cal F}={\cal F}(\bar{S},\bar{Y},\bar{T})\:.
\label{Fergdep}
\ee
As a function of these variables the normalized center of mass motion ${\cal F}$ is a universal function, i.e.\ independent of the precise microstructure of the polymers.  

The scaling variables have a simple interpretation. $\bar{T}$ measures time $t$ on the scale of the longest relaxation time $\sim n^{1+2\nu}$ of a free draining chain in the excluded volume limit: 
\be
\bar{T}\sim t/n^{1+2\nu}\:.
\label{Tsim}
\ee  
The variable $\bar{Y}$ is proportional to the ratio of the mean chain length $N$ of the background polymers and the chain length $n$ of the moving polymer:
\be
\bar{Y}\sim N/n\:.
\label{ysim}
\ee  
The overlap variable $\bar{S}$ requires a more detailed discussion: In the case of a binary semidilute polymer system, chain length and concentration occur only in the dimensionless combination $\bar{S}^{(B)}\sim c_p{R_g^{(B)d}}\sim cN^{d\nu-1}$ \cite{DeGe}. The overlap $\bar{S}^{(B)}$ describes the number of chains which are present in the volume occupied by one chain. In the dilute limit we have $\bar{S}^{(B)}\to 0$, while the semidilute limit is given by $\bar{S}^{(B)}\to\infty$. Now in our case we are interested in the properties of a tracer chain of different species and length, which moves in such a background. As observed in \cite{SchKa93,Nose86}, the appropriate overlap type variable at the symmetric excluded volume fixed point in such a ternary problem is $\bar{S}\sim c_pR_g^dN/n$ (which is called $s^{(S)}$ in \cite{SchKa93}). If we divide the background chains into blobs of length $n$ (assume $N\ge n$, see below), then $c_pN/n$ measures the number concentration of these blobs in the background. $\bar{S}$ is thus the number of such blobs which overlap in a volume $R_g^d\sim n^{d\nu}$ occupied by the moving chain. This variable ensures that it is irrelevant for the moving polymer, wether the background blobs belong to different chains or not, as it should be, since the moving chain at a given time does not know what happens on scales larger than its own extension. Note that $\bar{S}$ depends on $c$ and $n$ only:
\be
\bar{S}\sim cn^{d\nu-1}\:.
\label{Ssim}
\ee
The proportionality constants in eqs.~(\ref{Tsim})-(\ref{Ssim}) depend on the chemical microstructure of the polymers (cf.\ eqs.~(\ref{yquer})-(\ref{time})). In the framework of an RG analysis they can not be calculated reliably but they have to be taken as fit parameters instead.

Quite generally we expect the center of mass motion to become independent of $N$ if the background chains are much longer than the extension of the moving chain, $\bar{Y}\gg 1$. This conjecture becomes explicit in the one loop function $H^\ast_R$, eq.~(\ref{HR}):
\be
H_R^{\ast}(\bar{S},\bar{Y};\tau,z,s,y)=\frac{\bar{S}\bar{Y}D_p\Big({\tex\bar{Y}\frac{z}{\sqrt{2\tau}F(y,y+2s/\sqrt{2\tau},2/\sqrt{2\tau})}}\Big)}{1+\bar{S}\bar{Y}D_p\Big({\tex\bar{Y}\frac{z}{\sqrt{2\tau}F(y,y+2  s/\sqrt{2\tau},2/\sqrt{2\tau})}}\Big)}\:,
\nonumber\\
\label{HRfixu}
\ee
where the star signals, that we have set the couplings to their fixed point value ${u}^\ast$. Note further that we have made the dependence on the scaling variables $\bar{S},\bar{Y}$ explicit now. Due to the asymptotic behaviour of the Debyefunction $D_p(x) \stackrel{x\to\infty}{\sim}2/x$, cf.\ eq.~(\ref{Debasy}), we in the limit $\bar{Y}\to\infty$ obtain: 
\be
H_R^{\ast}(\bar{S},\infty;\tau,z,s,y)=\frac{2\bar{S}\sqrt{2{\tau}}F(y,y+2s/\sqrt{2\tau},2/\sqrt{2\tau})/z}{1+2\bar{S}\sqrt{2{\tau}}F(y,y+2s/\sqrt{2\tau},2/\sqrt{2\tau})/z}\:,
\nonumber\\
\label{HRfixasy}
\ee
such that the dependence on $\bar{Y}$ and thus on $N$ in one loop drops out indeed. More problematic is the opposite limit $\bar{Y}\to 0$, since in eq.~(\ref{HRfixu}) this limit would compete with the semidilute limit $\bar{S}\to\infty$. Now having in mind the discussion on the renormalized length scale at the end of the previous section, one may rightly ask wether the choice $n_R=1$ for $\bar{Y}\to 0$ is appropriate any longer. Since in the limit $N\ll n$ the length scale set by the moving chain is much greater than those of the background chains, one would argue for a choice $N_R=1$ instead. In eq.~(\ref{HRfixu}) this would lead to a replacement of $\bar{S}\bar{Y}$ by the overlap of the background $\bar{S}^{(B)}\sim cN^{d\nu-1}$. The limit $\bar{Y}\to 0$ then appears to be unproblematic, resulting in $\bar{S}^{(B)}D_p({\bar{Y}z/ {\sqrt{2{\tau}}F(y,y+2s/\sqrt{2\tau},2/\sqrt{2\tau})}})
\stackrel{\bar{Y}\to 0}{=}\bar{S}^{(B)}\tilde{p}''$, cf.\ eq.~(\ref{Debpol}). But this is not the whole story, since $H_R^{\ast}$ then would be independent of $\tau$, such that the integral $\int_0^{T_R} d\tau{\tau}^{-d/4}$ in eq.~(\ref{Rresren}) diverges in $d=4$ at the lower bound (a feature which also arises in the semidilute limit, see below). These problems occurring in a RG treatment of the limit of a very long chain in a system of moderately long chains are known to show up even in polymer statics \cite{DeGe,SchKa93,DeJa}. On a heuristic level it is clear that a semidilute system of very short background chains will eventually lead to the universality class of short range Gaussian random potentials. As a consequence of the new singularities for $\bar{Y}\to 0$, a precise analysis of this limit would however require a reconsideration of the renormalization of the couplings as well. We will not pursue this here and restrict ourselves in the following to $n\kgl N$, excluding the limit $\bar{Y}\to 0$ (see also appendix B).

Calculating the explicit scaling function we proceed with an $\epsilon$-expansion. In this context two things must be remembered: Firstly it is important to note that the renormalized function $H_R$ itself, eq.~(\ref{HR}), must not be expanded, since such an expansion would destroy the physical idea of screening. The structure of $H_R$ comes from a resummation in the treatment of many chain systems and thus has nothing to do with the critical behaviour treated by renormalization and $\epsilon$-expansion. Secondly, to extract anomalous power laws from an $\epsilon$-expansion we as usual have to exponentiate the perturbative result. Inserting eq.~(\ref{symfi}) into eq.~(\ref{Rresren}), the fixed point behaviour of the center of mass motion in first order $\epsilon$-expansion hence follows as:
\ba
\frac{{\cal F}(\bar{S},\bar{Y},\bar{T})}{\bar{T}}&=&\hspace{-0.0cm}\exp\Bigg\{-\frac{\epsilon}{64}\int_0^{\bar{T}} d\tau\, {\tau}^{-1} \left(1-\tau/\bar{T}\right) \int_0^\infty dz\,z^{2}e^{-z}\int_0^1 ds\nonumber\\
&&\qquad\quad \int_{-s/\sqrt{2\tau}}^{(1-s)/\sqrt{2\tau}}dy\,F(y,y+2s/\sqrt{2\tau},2/\sqrt{2\tau})^{-3}\nonumber\\
&&\qquad\quad H_R^{\ast}(\bar{S},\bar{Y};\tau,z,s,y)+{\cal O}({\epsilon}^2)\Bigg\}\:,
\label{Rresreneps}
\ea
where
$H_R^{\ast}$ is given in eq.~(\ref{HRfixu}).

\subsection{Diffusion constant}

We start with analyzing the diffusion constant $D$. In a way this is the simplest quantity, since here the dependence on time drops out due to the limit $t\to\infty$:
\be
D:=\lim_{t\to\infty}\frac{R^2(t)}{2dt}
\label{D}
\ee
We normalize this quantity by division by the diffusion constant of the free Rouse chain, $D_0=\gamma/n={\Gamma_0}$, eq.~(\ref{Gamma}). From eq.~(\ref{Rresreneps}) we obtain 
\ba
{\cal D}(\bar{S},\bar{Y})&:=&\frac{D}{D_0}=\lim_{\bar{T}\to\infty}\frac{{\cal F}(\bar{S},\bar{Y},\bar{T})}{\bar{T}}=\hspace{-0.0cm}\exp\Bigg\{-\frac{\epsilon}{64}\int_0^{\infty} d\tau\, {\tau}^{-1} \int_0^\infty dz\,z^{2}e^{-z}\nonumber\\
&&\hspace*{-2.5cm}\!\int_0^1\! ds\! \int_{-s/\sqrt{2\tau}}^{(1-s)/\sqrt{2\tau}}\!\!dy \,F(y,y+2s/\sqrt{2\tau},2/\sqrt{2\tau})^{-3}\, H_R^{\ast}(\bar{S},\bar{Y};\tau,z,s,y)+{\cal O}({\epsilon}^2)\Bigg\}.
\nonumber\\
\label{Dres}
\ea
In the dilute limit, $H_R^{\ast}\sim\bar{S}\to 0$, cf.\ eq.~(\ref{HRfixu}), we of course recover Rouse behaviour,
\be
\lim_{\bar{S}\to 0}{\cal D}(\bar{S},\bar{Y})=1\:.
\label{Ddil}
\ee
Recall that the intrachain excluded volume interaction $u$, which is still present also for $\bar{S}\to 0$, does not affect the center of mass motion as long as the disorder vanishes. For small $\bar{S}$ eq.~(\ref{Dres}) can be evaluated as a virial expansion in powers of $\bar{S}$.

Much more interesting is the limit of strongly overlapping chains, $\bar{S}\gg 1$. Since $\lim_{\bar{S}\to\infty}H_R^{\ast}=1$ (cf.\ eq.~(\ref{HRfixu})), a glance at eq.~(\ref{Dres}) immediately reveals that this limit if taken naively would produce a divergence due to the integral $\int_0 d\tau{\tau}^{-1}$. This was to be expected from the discussion of sect.\ 4.2, since here we obtain a structure similar to the limit $c\to\infty$ in the unrenormalized result, which is singular in $d=4$. The resulting $\epsilon$-pole in \cite{utestat} induced a nontrivial renormalization of $\Gamma_0$. Here it translates into a logarithmic divergence in $\bar{S}$. To extract this divergence we exploit the asymptotic behaviour of the integrand in eq.~(\ref{Dres}) for small $\tau$ (cf.\ eqs.~(\ref{diver1}),(\ref{Debasy})):
\be
H_R^{\ast}(\bar{S},\bar{Y};\tau,z,s,y)\stackrel{\tau\to 0}{=}\frac{2\bar{S}\sqrt{2{\tau}}f(y)/z}{1+2\bar{S}\sqrt{2{\tau}}f(y)/z}\:.
\label{HRfixst}
\ee
With this asymptotic form we can perform the dangerous part of the $\tau$ integral near the lower bound in eq.~(\ref{Dres}) and find the desired logarithmic divergence:
\be
\int_0^{1} d\tau\, {\tau}^{-1} \int_0^\infty dz\,z^{2}e^{-z}\int_0^1 ds\int_{-\infty}^{\infty}dy\,f(y)^{-3}\frac{2\bar{S}\sqrt{2{\tau}}f(y)/z}{1+2\bar{S}\sqrt{2{\tau}}f(y)/z}\stackrel{\bar{S}\to\infty}{\sim}8I\ln {\bar{S}}\:,
\label{Dasyzwi}
\ee
where $I=3.587\ldots$ is given in eq.~(\ref{I}). Isolating this divergent term in eq.~(\ref{Dres}) we obtain
\be
{\cal D}(\bar{S},\bar{Y})={\bar{S}}^{-\epsilon I/8+{\cal O}({\epsilon}^2)}{\bar{\cal R}}(\bar{S},\bar{Y})\:,
\label{Dfinasy}
\ee
where the residual term ${\bar{\cal R}}$ stays finite in the semidilute limit: \be
\lim_{\bar{S}\to\infty}{\bar{\cal R}}(\bar{S},\bar{Y})=\exp(.56\epsilon+{\cal O}({\epsilon}^2))\stackrel{d=3}{=}1.75\:.
\label{barRsem}
\ee
Note that the right hand side of eq.~(\ref{barRsem}) is independent of $\bar{Y}$ and of polydispersity. This was to be expected, since in the limit of strongly overlapping background chains, the moving polymer effectively feels uncorrelated concentration blobs of the background only. The chain length of the background polymers completely drops out. The result eq.~(\ref{Dfinasy}) shows that the diffusion coefficient in the semidilute limit obeys a power law. With eqs.~(\ref{nu}),(\ref{Ssim}) and $d=4-\epsilon$ we find: 
\be
D=D_0\,{\cal D}(\bar{S},\bar{Y})\stackrel{\bar{S}\gg 1}{\sim}c^{-\epsilon I/8+{\cal O}({\epsilon}^2)}n^{-1-\epsilon I/8+{\cal O}({\epsilon}^2)}\:.
\label{Dpower0}
\ee
In leading nontrivial order of $\epsilon$-expansion this yields in three dimensions ($\epsilon=1$) 
\be
D\sim c^{-.45}n^{-1.45}\:.
\label{Dpower}
\ee
Compared to Rouse behaviour, $D_0\sim n^{-1}$, we obtain a slowing down, which however is not as strong as in the reptation picture, $D_{rept.}\sim n^{-2}$ \cite{DeGerept}. One concludes that the hindering due the disordered excluded volume interactions of a frozen semidilute background is not enough to obtain a scaling behaviour as if true topological constraints were present. Also the dependence on the monomer concentration is comparatively weak. Reptation, combined with simple scaling considerations would yield a much stronger slowing down with increasing concentration: $D_{rept.}\sim c^{-1.31}n^{-2}$. The power -1.31 follows from the requirements that both $D_{rept.}/D_0$ must be a function of $\bar{S}$ only and $D_{rept.}\sim n^{-2}$. In the reptation case such a scaling argument however is dubious, since it neglects the fact that topological constraints generally define a new length scale (entanglement length), which is different from $\xi_c$, the size of concentration blobs \cite{LoRo}. We close this short discussion of power law behaviour with the remark that the numerical values for the exponents we have obtained should not be taken too serious, since they are based on a first order $\epsilon$-expansion.

In fig.\ \ref{Dnew} we have plotted the full dependence of the normalized diffusion coefficient on $\bar{S}$ according to eq.~(\ref{Dres}). We have chosen a monodisperse ensemble, i.e.\ all chains of the background have the same length $N$. This simply amounts to a replacement of $D_p$, eq.~(\ref{Debpol}), by the standard Debye function $D(x)=2/x^2(e^{-x}-1+x)$. Regarding the chain lengths we consider the two cases $\bar{Y}=1$, which holds if $n=N$ and the moving chain is chemically identical to the background polymers ($B_1=B_2$), resp.\ $\bar{Y}=\infty$, which describes the case of a long tracer chain that moves in a frozen background of much longer chains, $N\gg n$. Note that the curves are universal, i.e.\ independent of microstructure effects. In the semidilute limit the diffusion constant becomes independent of $\bar{Y}$: Both curves obviously coincide for $\bar{S}\gg 1$ and follow the asymptotic form $1.75\,{\bar{S}}^{-.45}$ of eqs.~(\ref{Dfinasy}),(\ref{barRsem}). Quite generally the dependence on $\bar Y$ is very small, a property also known from static quantities \cite{SchKa93}. Note further that ${\cal D}(\bar{S},\infty)<{\cal D}(\bar{S},1)$. The reason is that keeping $c,n$ and thus $\bar{S}$ fixed, an increasing $N$ means that the chain concentration of the background $c_p=c/N$ decreases. With rising $\bar{Y}$ the constant number of background segments hence belong to fewer and fewer chains. But the more the background segments are correlated due to chain connectivity, the more effective is the hindering: The diffusion constant drops with increasing $\bar{Y}$.

\subsection{Full time dependence}
As explained in the introduction we in a semidilute background expect three time regimes:\\
(i) For very short times $\bar{T}\kgl {\bar{S}}^{-2}$ the moving polymer does not notice its environment, i.e.\ we expect essentially free diffusion. In this ultra-short time regime the polymer moves a distance much less than a correlation length of the background: $R(t)\ll {\xi}^{(B)}$. In \cite{utestat} this regime does not exist, since the correlation length is microscopic: ${\xi}^{(B)}\to 0$.\\ 
(ii) What follows is a short time regime $\bar{T}\kgl 1$ but $\bar{T}{\bar{S}}^{2}\gg 1$, where the polymer interacts with the background and slows down. The motion then is subdiffusive.\\
(iii) In the long time regime $\bar{T}\gg 1$, i.e.\ $R(t)\gg R_g$, we obtain a diffusive motion of the polymer coil as a whole, but with reduced diffusion coefficient. 

This supposed general behaviour can be confirmed explicitly in our one loop result eq.~(\ref{Rresreneps}). Note that in this section we always assume strongly overlapping chains: $\bar{S}\gg 1$.\\
{\bf (i) Ultra-short time behaviour:}\\ 
For vanishing $\bar{T}$ we have    
\be
\lim_{\bar{T}\to 0}\frac{{\cal F}(\bar{S},\bar{Y},\bar{T})}{\bar{T}}=1\:,
\label{Rresqulim}
\ee
which means that there must be an ultra-short time regime where the correction to the leading behaviour eq.~(\ref{Rresqulim}) remains small. Indeed, for times $\bar{T}\kgl {\bar{S}}^{-2}$, even for very large $\bar{S}$ no divergences in the time integral show up. This can be seen from eq.~(\ref{HRfixst}), which reveals that as long as $\sqrt{{\tau}}\bar{S}$ is small, the one loop correction in eq.~(\ref{Rresreneps}) is small.\\ 
{\bf (ii) Short time behaviour}\\ 
Leaving the ultra-short time regime we analogous to eq.~(\ref{Dasyzwi}) have to extract the asymptotic behaviour. Using eq.~(\ref{HRfixst}), we find  
\be
\int_0^{\bar T}\!\! d\tau\, {\tau}^{-1}\! \int_0^\infty\!\! dz\,z^{2}e^{-z}\int_0^1 \!\!ds\int_{-\infty}^{\infty}\!\!dyf(y)^{-3}\frac{2\bar{S}\sqrt{2{\tau}}f(y)/z}{1+2\bar{S}\sqrt{2{\tau}}f(y)/z}
\stackrel{\bar{S}\to\infty}{\sim}8I\ln\! \sqrt{\bar T}{\bar{S}}\,,
\label{Fasyzwi}
\ee
so that
\be
\frac{{\cal F}(\bar{S},\bar{Y},\bar{T})}{\bar{T}}= {\left({\sqrt{\bar{T}}\bar{S}}\right)}^{-\epsilon I/8+{\cal O}({\epsilon}^2)}{\cal R}(\bar{S},\bar{Y},\bar{T})\:.
\label{Ffinasy}
\ee
In the semidilute limit ${\cal R}(\bar{S},\bar{Y},\bar{T})$ exists and is independent of $N$ and polydispersity, as it should be (cf.\ eq.~(\ref{barRsem})):
\be
\lim_{\bar{S}\to\infty}{\cal R}(\bar{S},\bar{Y},\bar{T})=:\hat{{\cal R}}(\bar{T})\:.
\label{Rsemidil}
\ee 
Note that a time dependence still remains, such that eq.~(\ref{Ffinasy}) even in the semidilute limit does not yield a pure power law with respect to time. However, taking in eq.~(\ref{Rsemidil}) the limit $\bar{T}\to 0$, we find:
\be
\hat{{\cal R}}(0)=\exp\left(.30\epsilon+{\cal O}({\epsilon}^2)\right)\stackrel{d=3}{=}1.35\:.
\label{RsemT}
\ee
This implies that for large overlap there exists an intermediate regime
\be
1\ll \bar{T}{\bar{S}}^{2}, \bar{T}\ll 1 
\label{shotimereg}
\ee 
i.e. $\mbox{const.}\,c^{-2}n^{3+2\nu(1-d)}\ll t\ll\mbox{const.}\,n^{1+2\nu}$, where $R^2(t)$ approaches a power law:
\be
R^2(t)\sim c^{-\epsilon I/8+{\cal O}({\epsilon}^2)}n^{-1+{\cal O}({\epsilon}^2)}t^{1-\epsilon I/16+{\cal O}({\epsilon}^2)}\stackrel{d=3}{\sim} c^{-.45}\,n^{-1}\,t^{.78}\:.
\label{Rfinasy}
\ee
This time dependence again lies between Rouse behaviour, $R^2(t)\sim t$, and the short time reptation result, $R^2(t)\sim t^{.5}$. The trivial dependence on chain length $\sim n^{-1}$ should survive in all orders of $\epsilon$-expansion, if we can think of the chain as being composed of $n/n_{\xi}$ time blobs of size $\xi_t$, which move independently from each other \cite{EbBaSch}. The latter condition is fulfilled for $1\ll \bar{T}{\bar{S}}^{2}$, which implies $\xi_c\ll\xi_t$, so that the semidilute background can not build up correlations between the blobs.\\
{\bf (iii) Long time behaviour}\\ 
To describe the long time behaviour $\bar{T}\gg 1$ we have to do the same considerations as for the diffusion constant, eqs.~(\ref{Dasyzwi})-(\ref{barRsem}). Eq.~(\ref{Ffinasy}) is not appropriate since $\lim_{\bar{T}\to\infty}{\cal R}(\bar{S},\bar{Y},\bar{T})$ does not exist. As above we obtain:
\be
\frac{{\cal F}(\bar{S},\bar{Y},\bar{T})}{\bar{T}}= {\bar{S}}^{-\epsilon I/8+{\cal O}({\epsilon}^2)}{\tilde{\cal R}}(\bar{S},\bar{Y},\bar{T})\:,
\label{Fasylt}
\ee
Note that $\lim_{\bar{T}\to\infty}{\tilde{\cal R}}(\bar{S},\bar{Y},\bar{T}) = {\bar{\cal R}}(\bar{S},\bar{Y})$ of eq.~(\ref{Dfinasy}), such that the long time behaviour is purely diffusive:
\be
R^2(t)\stackrel{\bar{T}\gg 1}{=}2dD_0t{\bar{S}}^{-\epsilon I/8+{\cal O}({\epsilon}^2)}{\bar{\cal R}}(\bar{S},\bar{Y})=2dDt\:.
\label{diffusive}
\ee
The diffusion constant $D$ has been discussed in sect.\ 5.1.

In fig.\ \ref{Fnew} we have plotted the universal time dependence of ${\cal F}=R^2(t)/2dn_Rl_R^2$ according to eq.~(\ref{Rresreneps}) for three values $\bar{S}=10,100,1000$ of the overlap. As before we restrict ourselves to a monodisperse ensemble. Since the dependence on $\bar{Y}$ is expected to be weak (cf.\ fig.\ \ref{Dnew}), we here show the case of identical chains only: $\bar{Y}=1$. All curves start out with a Rouse like ultra-short time behaviour, then bend down to anomalous diffusion and end up with normal diffusion again, $R^2(t)\sim t$, but with reduced diffusion coefficient $D$ as compared to the Rouse value $D_0$. For large overlap one may identify an effective short time power law regime ($R^2(t)\sim t^{.8}$ for $\bar{S}=1000$), the value of the power depending on the overlap $\bar{S}$ however. Only in the asymptotic limit $\bar{S}\to\infty$, $\bar{T}\to 0$, with $\bar{T}{\bar{S}}^2\to\infty$, we obtain the true universal short time power law of eq.~(\ref{Rfinasy}): $R^2(t)\sim t^{.78}$.

\section{Conclusions}

We have presented a systematic perturbative approach to polymer dynamics in a quenched many chain background on the basis of standard RG methods. The model describes the hindering due to the excluded volume repulsion of a test chain moving in a frozen semidilute polymer background. Topological constraints are not included. We carried through an explicit one loop calculation for the center of mass motion and identified the scaling variables. In the semidilute limit we have found power law behaviour with new exponents. They exhibit a slowing down of the polymer motion with increasing overlap which however is not as pronounced as in the reptation picture. 

Trying to compare our results with experimental measurements we are confronted with a very sparse amount of data available. Clearly real experiments always involve entanglement effects. On the other hand one might argue that for low concentrations $c$ (but very long chains $n$, such that we still work with great overlap ${\bar S}\sim cn^{d\nu-1}$), entanglement might not be that important, at least for short times. A further problem is that in real experiments we have an annealed random background, while our theory is set up for a quenched average. Only in case of a gel, where the crosslinks fix the network to a large extend, or, at least partially, if we deal with background chains that are much longer than the test chain one might advocate for a separation of time scales, such that the random background may be taken as effectively frozen as compared to the tracer polymer. Nevertheless a quantitative comparison between our theory and real experiments does not seem to be very promising.  

We therefore must rely on computer experiments. The problem here is that most of the work we are aware of focuses on the simulation of polymer melts, aiming at a verification of the reptation predictions. This means fairly concentrated systems and in most cases annealed averages, i.e.\ all the chains move simultaneously. Kremer \cite{Krem83} in his early work performed one simulation with one mobile test chain in a frozen environment of $M=6$ other chains, each of length $n=200$, but he did not vary chain length and concentration, making a quantitative comparison impossible. The one simulation he carried through moreover belongs to the concentrated regime, where entanglements are not negligible. The same objections hold for many other simulations, a review can be found in \cite{LoRo}.  

For a real test of the theory one would need a quenched polymer background at very low concentration but with very long chains, such that we achieve a high enough overlap ${\bar S}\gg 1$ to observe the asymptotic scaling behaviour we predicted. Simulations in this direction would be most welcome. For a more comprehensive insight into the dynamics of our model we furthermore started to calculate the time dependent segment-segment correlation function, which exhibits a much richer dynamic behaviour. 

\section*{Acknowledgement}
I would like to thank Prof.\ L.\ Sch\"afer for many helpful discussions and for a careful reading of the manuscript. This work has been supported by the Deutsche Forschungsgemeinschaft, SFB `Unordnung und gro{\ss}e Fluktuationen'.

\begin{appendix}
   \renewcommand{\thesection}{}
   \renewcommand{\thesubsection}{\Alph{subsection}}
   \renewcommand{\theequation}{\Alph{subsection}.\arabic{equation}}
   \renewcommand{\thefigure}{\Alph{subsection}.\arabic{figure}}
\section{Appendix}

\setcounter{equation}{0}
\subsection{Quenched vs.\ annealed in polymer statics}
As announced in the main text, we here will shortly rephrase the self-averaging argument of \cite{WuHui,utemacro}:\\
Divide the space into boxes of length $a$. A state of the chain $\{{\bf r}\}$ could equally be described by the starting point ${\bf r}_{1}$ and the internal configuration $\{{\bf c}\}$, where ${\bf c}_i = {\bf r}_i-{\bf r}_{i-1}\,(i=2,\ldots,n)$. Splitting the integration over ${\bf r}_{1}$ in an integration over a box volume and the sum over all boxes, such that ${\bf r}_{1}={\bf k}+{\bf s}$, ${\bf k}\in a\cdot{\Bbb{Z}}^d$, we can write eq.~(\ref{Zque}) as
\be
{\cal Z}_M[\{{\bf r}^{(B)}\}]=\int{\cal D}\{{\bf c}\}\, \int_0^a \frac{d^ds}{(4 \pi l^2)^{d/2}}\sum_{{\bf k}\in a\cdot{\Bbb{Z}}^d}\e^{\tex -{\cal H}_M}\:. 
\label{Zquespl}
\ee
Now concentrate on one fixed internal configuration $s$ and $\{{\bf c}\}$ of the partition function in eq.~(\ref{Zquespl}) (see fig.\ \ref{figboxes}). Then with respect to disorder average all the summands (=boxes in fig.\ \ref{figboxes}) in the infinite ($\Omega=L^d\to\infty$) sum $\sum_{{\bf k}\in a\cdot{\Bbb{Z}}^d}$ are statistically independent, if all segments in one box are separated more than a background correlation length from all segments of all other boxes. This is guaranteed by choosing the box size to be of the order $a\ggl{\cal O}(R_g)+{\cal O}(R_g^{(B)})$, taking into account that $\xi^{(B)}\le R_g^{(B)}$. The law of large numbers then yields that with probability one $\sum_{{\bf k}\in a\cdot{\Bbb{Z}}^d} e^{ -{\cal H}_M}=\overline{\sum_{{\bf k}\in a\cdot{\Bbb{Z}}^d} e^{ -{\cal H}_M}}$ holds. 

To illustrate this argument we in \cite{disseng} carried out an explicit one loop calculation for the two point density correlation function between the tracer chain and the background chains. This is a static observable, that depends on the interchain coupling already in tree approximation. We on the one hand calculated the annealed average via diagrammatic perturbation theory of ternary solutions, on the other hand we calculated the quenched average via the dynamic functional. Note that in this case we used the dynamic functional to calculate a static quantity, since in dynamics the quenched average is carried out easily (no problems with the partition function in the denominator as in quenched static averages). The lengthy calculations within the framework of two completely different calculation schemes yielded two results, which for $\Omega=L^d\to\infty$ exactly coincide indeed.

\setcounter{equation}{0}
\subsection{A brief discussion of other fixed points}
In the main text we exclusively evaluated our theory at the symmetric excluded volume fixed point, where all three couplings $u$,$u^{(B)}$ and $v$ coincide and are equal to the binary fixed point value $u^{\ast}\neq 0$. One may ask what comes out if we would have dealt with Gaussian chains, where one or both couplings $u$ or $u^{(B)}$ vanish. To this end one might set e.g.\  $u^{(B)}=0$ in our one loop result eq.~(\ref{Rresren}). A short glance at eq.~(\ref{HR}) immediately reveals, that in this case the limit of strongly overlapping chains $c_R\sim{\bar S}\gg 1$ can not be taken, since the one loop correction term simply diverges $\sim{\bar S}$. (We always assume $v\neq 0$, since otherwise the test chain is decoupled from the background.) 

This problem is closely related to the breakdown of sreening in ternary semidilute solutions. Generally speaking the intrachain coupling $u$ of a test polymer in a solution of other chains becomes screened, $u\to {\tilde u}$, where the renormalized screened interaction ${\tilde u}$ reads \cite{SchKa93}:
\be
{\tilde u}(z)=u\:\frac{1+\left(1-\frac{\tex v^2}{\tex uu^{(B)}}\right)u^{(B)}c_RN_RD_p\left(\frac{N_R}{n_R}\,z\right)}{1+u^{(B)}c_RN_RD_p\left(\frac{N_R}{n_R}\,z\right)}\:. 
\label{uscree}
\ee
Here $z$ denotes a momentum type variable. Being interested in the large scale properties we can set $z=0$. At the symmetric fixed point we then obtain complete screening in the semidilute limit: ${\tilde u}=0$ for $c_R\sim {\bar S}\to\infty$. (We however note that  ${\tilde u}=u$ in the limit $N_R\to 0$: Short background chains can not screen the interactions within a long test chain, a property wich refers to the fact that we did not analyze the limit ${\bar Y}\sim N/n\to 0$ in the main text.) On the other hand, in case of either Gaussian background chains, $u^{(B)}=0$, or a Gaussian test chain, $u=0$, or both, we obtain ${\tilde u}<0$ for large overlap: The test chain collapses.

The physical picture behind that analysis is that a polymer solution of two incompatible species demixes. Incompatibility is signaled by $v^2>uu^{(B)}$ and eventually leads to a phase separated equilibrium state: A test chain is immediately driven out of the background chains. Equlibrium dynamics with a tracer chain moving through a semidilute background is only possible for a compatible system. With such a system we are dealing in the main text.

In conclusion we note that the center of mass motion in one loop is independent of $u$, such that ${\tilde u}$ does not show up in eq.~(\ref{Rresren}). For other dynamic observables like segment correlations we actually do find a contribution similar to ${\tilde u}$ in a one loop calculation.

\end{appendix}
\bibliography{literatur}
\bibliographystyle{myunsrt}
\newpage

\setcounter{figure}{1}
\newcounter{fig}
\newcommand{\alphfig}{\setcounter{fig}{\value{figure}}%
\setcounter{figure}{0}%
\renewcommand{\thefigure}{%
\mbox{\arabic{fig}\alph{figure}}}}%
\newcommand{\resetfig}{\stepcounter{fig}%
\setcounter{figure}{\value{fig}}%
\renewcommand{\thefigure}{\arabic{figure}}}

\vspace*{-3.0cm}
\section* {Figures:}
\vspace*{.5cm}
\setcounter{figure}{0}

\begin{figure}[hbt]
\hspace*{.7cm}\epsfxsize12.0cm \parbox{12.0cm}{\epsfbox{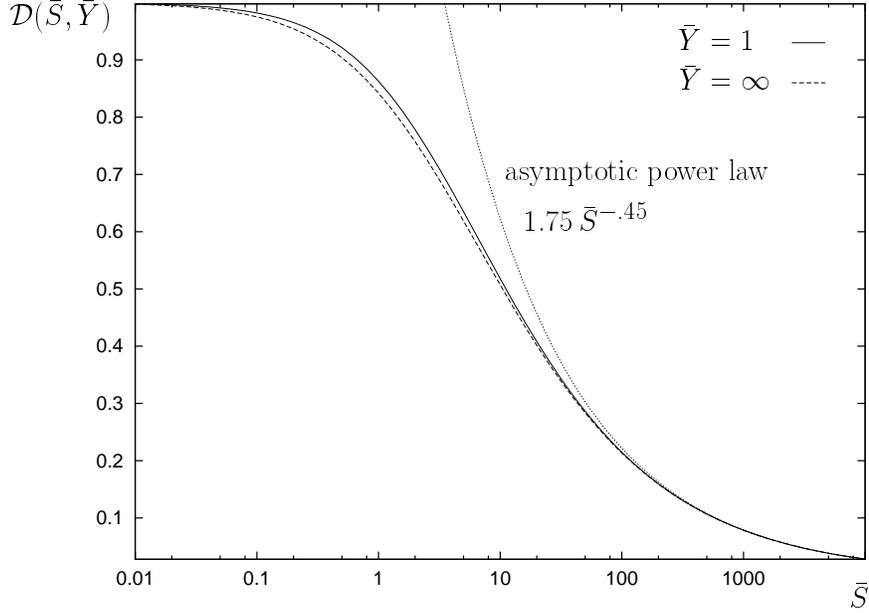}}
\caption[Fig.1]{\label{Dnew} Normalized diffusion constant ${\cal D}(\bar{S},\bar{Y})=D/D_0\sim nD$ as a function of the overlap $\bar{S}\sim cn^{d\nu-1}$, plotted for a monodisperse background and $\bar{Y}=1$ resp.\ $\bar{Y}=\infty$ (dashed). Dotted: The asymptotic power law behaviour according to eqs.~(\ref{Dfinasy}),(\ref{barRsem}).}
\end{figure}

\begin{figure}[hbt]
\hspace*{.7cm}\epsfxsize12.0cm \parbox{12.0cm}{\epsfbox{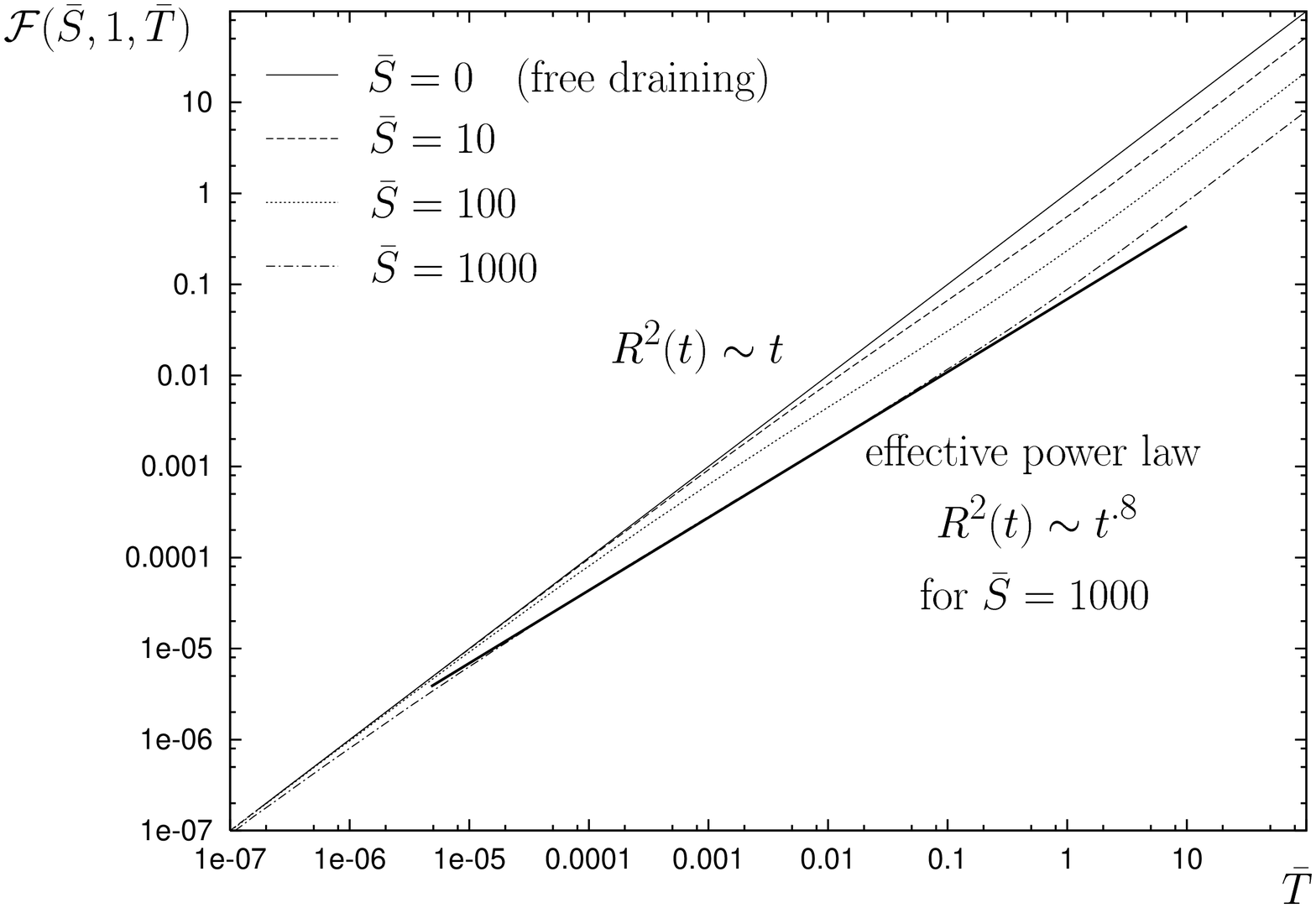}}
\caption[Fig.2]{\label{Fnew} Full time dependence of the normalized center of mass motion ${\cal F}\sim R^2(t)/n^{2\nu}$ as a function of normalized time $\bar{T}\sim t/n^{1+2\nu}$ for three values of the overlap ($\bar{S}=10,100,1000$) together with the free case ($\bar{S}=0$). The quenched background is monodisperse and $\bar{Y}=1$.}
\end{figure}

\begin{figure}[hbt]
\epsfxsize12.0cm \parbox{12.0cm}{\epsfbox{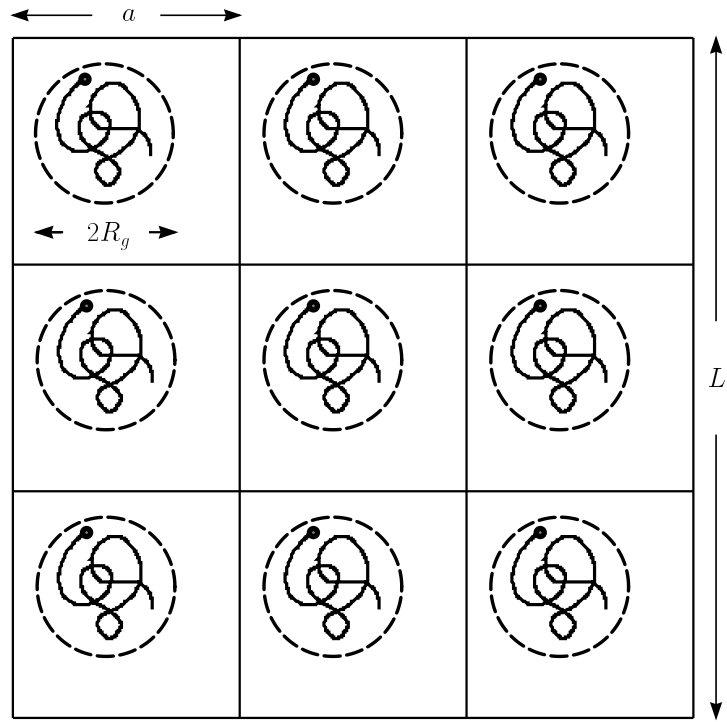}}
\caption[Fig.3]{\label{figboxes} Illustration of the self-averaging property of the partition function of a single chain in disorder.}
\end{figure}

\end{document}